\documentclass[5p,times]{elsarticle}
\usepackage{blindtext}
\usepackage{graphicx}
\usepackage{multirow}
\usepackage{booktabs}
\usepackage{comment}
\usepackage{xcolor}
\usepackage{amssymb}
\usepackage{mathtools}
\usepackage{hyperref}
\usepackage{tikz}
\usepackage{pgfplots}
\usepackage{color, colortbl}
\usepackage{bbm}
\usepackage{algorithm}
\usepackage{listings}

\usepackage{url}

\begin{document}

\begin{frontmatter}

\title{Segmentation Ability Map: Interpret deep features for medical image segmentation}

\author[1]{Sheng He\corref{cor1}\fnref{cof}}
\author[2]{Yanfang Feng\fnref{cof}}
\author[1]{P. Ellen Grant}
\author[1]{Yangming Ou\corref{cor1}}

\fntext[cof]{These authors contributed equally to this work.}
\cortext[cor1]{Corresponding author: heshengxgd@gmail.com, yangming.ou@childrens.harvard.edu}

\address[1]{Boston Children's Hospital and Harvard Medical School, 300 Longwood Ave., Boston, MA, USA}
\address[2]{Massachusetts General Hospital and Harvard Medical School, 55 Fruit St., Boston, MA, USA.}

\begin{abstract}
Deep convolutional neural networks (CNNs) have been widely used for medical image segmentation.
In most studies, only the output layer is exploited to compute the final segmentation results and the hidden representations of the deep learned features have not been well understood. 
In this paper, we propose a prototype segmentation (ProtoSeg) method to compute a binary segmentation map based on deep features.
We measure the segmentation abilities of the features by computing the Dice between the feature segmentation map and ground-truth, 
named as the segmentation ability score (SA score for short).
The corresponding SA score can quantify the segmentation abilities of deep features in different layers and units to understand the deep neural networks for segmentation.
In addition, our method can provide a mean SA score which can give a performance estimation of the output on the test images without ground-truth.
Finally, we use the proposed ProtoSeg method to compute the segmentation map directly on input images to further understand the segmentation ability of each input image.
Results are presented on segmenting tumors in brain MRI, lesions in skin images, COVID-related abnormality in CT images, prostate segmentation in abdominal MRI, and pancreatic mass segmentation in CT images.
Our method can provide new insights for interpreting and explainable AI systems for medical image segmentation.
 Our code is available on: \url{https://github.com/shengfly/ProtoSeg}.

\end{abstract}


\end{frontmatter}

\section{Introduction}
\label{sec:intro}

Medical image segmentation aims to train a machine learning model (such as the deep neural network~\citep{ronneberger2015u}) to learn the features of target objects from expert-annotations and apply it to test images.
Deep convolutional neural networks are popular for medical image segmentation~\citep{milletari2016v,zhou2019unetplusplus,wang2020pairwise,van2020hooknet} and a typical neural network contains an input layer which receives the medical images as inputs, several convolutional layers which extract deep features and an output layer which provides the final segmentation results.
Deep features are the activation maps from the hidden convolutional layers, indicating the locations and feature strength of hidden units (or filters) in deep neural networks~\citep{mou2020cs2}.
The neural network can decompose the input image into different deep features over layers toward the final output layer to separate the object and 
background pixels/voxels (such as the negative class or the non-annotated tissue on medical images) in the input image.
The segmentation ability measures the ability of a feature map to separate the object and background at each pixel/voxel.
The intensity values of the object and background pixels/voxels on the input images are usually not separable which has a low segmentation ability.
However, the intensity values of the segmentation map from the last layer of the neural network with Sigmoid/Softmax operation are separable which has a high segmentation ability.
Thus, there is a transition of the segmentation ability from the input image to the output of the neural network.
One question is raised: \textbf{where (or from which layer) are the features of the object and background regions separable or how the segmentation ability transits from the input image to the output of the neural network}?

Answering this question requires us to quantify and visualize the object regions on deep features in hidden layers of the neural network.
Quantitatively measuring the segmentation ability of the hidden deep features is useful for understanding or interpreting the neural network for segmentation.
However, only deep features of the output layer can be quantitatively interpreted because it represents the segmentation result which can be evaluated based on the ground-truth.
The deep features from the hidden convolutional layers of the segmentation neural network are not well understood.
The aim of this paper is to develop a tool (1) for data scientists or model developers to understand the neural networks and provide some insights for improving or developing the segmentation neural networks and (2) for end users to understand how U-Net makes the segmentation by different layers to enable human users to understand and appropriately trust the transparent machine learning model~\citep{arrieta2020explainable}.

One indirect method for understanding the hidden deep features is to use the attention mechanism to highlight salient deep features for the target task~\citep{schlemper2019attention,liu2020attention} by visualizing the learned attention~\citep{gu2020net}.
However, attention cannot be used for quantifying the importance of each deep feature or understanding how the decision is obtained on each pixel/voxel.
In addition, it can only highlight the salient regions on these layers where attention is used.

To solve this problem, we propose a method to exploit the rich information contained in intermediate deep features on neural networks for segmentation.
Ideally, the distance among features from the pixels/voxels in the same class (object/background) should be small while the distance among features from the pixel/voxels in different classes should be high.
Based on this assumption, we propose a simple \textit{prototype segmentation} (ProtoSeg for short) method to compute a binary segmentation map on deep features and then measure the segmentation ability by comparing it to the ground-truth.
Computing a binary segmentation is very useful for understanding the deep features in neural networks~\citep{zhou2018interpreting}. 
Given a deep feature, the prototypes of the object and background regions are computed as the mean of the feature values guided by the initial segmentation from the output of the neural network.
After that, all pixels/voxels on the feature map can be segmented based on the prototypes, yielding a feature segmentation map, named \textbf{segmentation ability map} (or SAM for short).
The segmentation ability can be measured by any metric used for segmentation evaluation between the SAM and the ground-truth.
In this paper, we use the well-established Dice score to measure the segmentation ability of SAM, named segmentation ability score (SA score).
A high SA score means a high segmentation ability which indicates the SAM of the deep feature close to the ground-truth.

The proposed ProtoSeg method is a plug-and-play module and is efficient without parameters.
It has three advantages.
First, it aims to understand the segmentation ability of deep features in different layers of the neural network.
A powerful deep feature should have a high segmentation ability and output a segmentation map that is close to the ground-truth, indicating that features on object regions are different from features on normal ones.
Second, the proposed ProtoSeg is differentiable, which can be used in different ways to measure the segmentation ability of different deep features: offline ProtoSeg and online ProtoSeg.
For offline ProtoSeg, it can compute the SAM on any deep feature extracted on the trained neural network for interpretation.
For online ProtoSeg, the segmentation ability computed by the ProtoSeg is differentiable and can be jointly trained to increase the segmentation ability of hidden features.
Third, the SAM can be used to estimate the confidence scores of the network's output from different input images by 
computing the mean SA score of the neurons on the last two layers.

 The main contributions of this paper are summarized as follows:
\begin{itemize}
    \item We propose a plug-and-play ProtoSeg method to compute the SAM on any deep feature map. The ProtoSeg method is parameter-free, simple, differentiable and computationally efficient.
    \item A SA score can be computed based on the binarized SAM map to interpret the deep features of the neural network.
    \item The proposed ProtoSeg can be used in different ways: offline and online. For online ProtoSeg, the SA score can be used in the training loss to improve the segmentation ability without decreasing the final accuracy.
    \item We apply the proposed method to understand the well-established U-Net on five different datasets, providing some insights for understanding the U-Net for pixel/voxel-wise segmentation.
\end{itemize}

\section{Background and motivation}
Interpreting deep networks is important for explainable machine learning and medical image segmentation.
Attention is usually used for understanding deep features which consists of the channel attention~\citep{hu2018squeeze} for highlighting the channels in each layer, spatial attention~\citep{schlemper2019attention} for visualizing the salient locations and the hybrid of spatial and channel attention~\citep{lei2020self}.
A comprehensive attention-based neural network is proposed in~\cite{gu2020net} for explainable medical image segmentation, including spatial, channel and scale attentions.
The limitation of using attention for interpretation is that it can only visualize the salient regions computed by the attention instead of quantifying the segmentation ability of the deep features.
Features with high attention may not necessarily be most useful for separating objects from background regions.
In addition, the salient regions can only be computed on the layers where attention is applied.
Thus, the limitation is that we cannot quantitatively compare attentions among different layers in the deep neural networks to understand the segmentation ability of the whole neural network.

For image recognition, deep features can be interpreted by the network dissection~\citep{zhou2018interpreting} or concept whitening~\citep{chen2020concept} which aim to understand the relationship between the activation maps (deep features) of every internal convolutional unit and the concepts defined by humans.
Another method of interpreting the deep features is the class activation map (CAM) which can be computed directly on the last convolutional layer~\citep{zhou2016learning} or any layers using gradients~\citep{selvaraju2017grad}.

\begin{figure}[!t]
    \centering
    \includegraphics[width=0.5\textwidth]{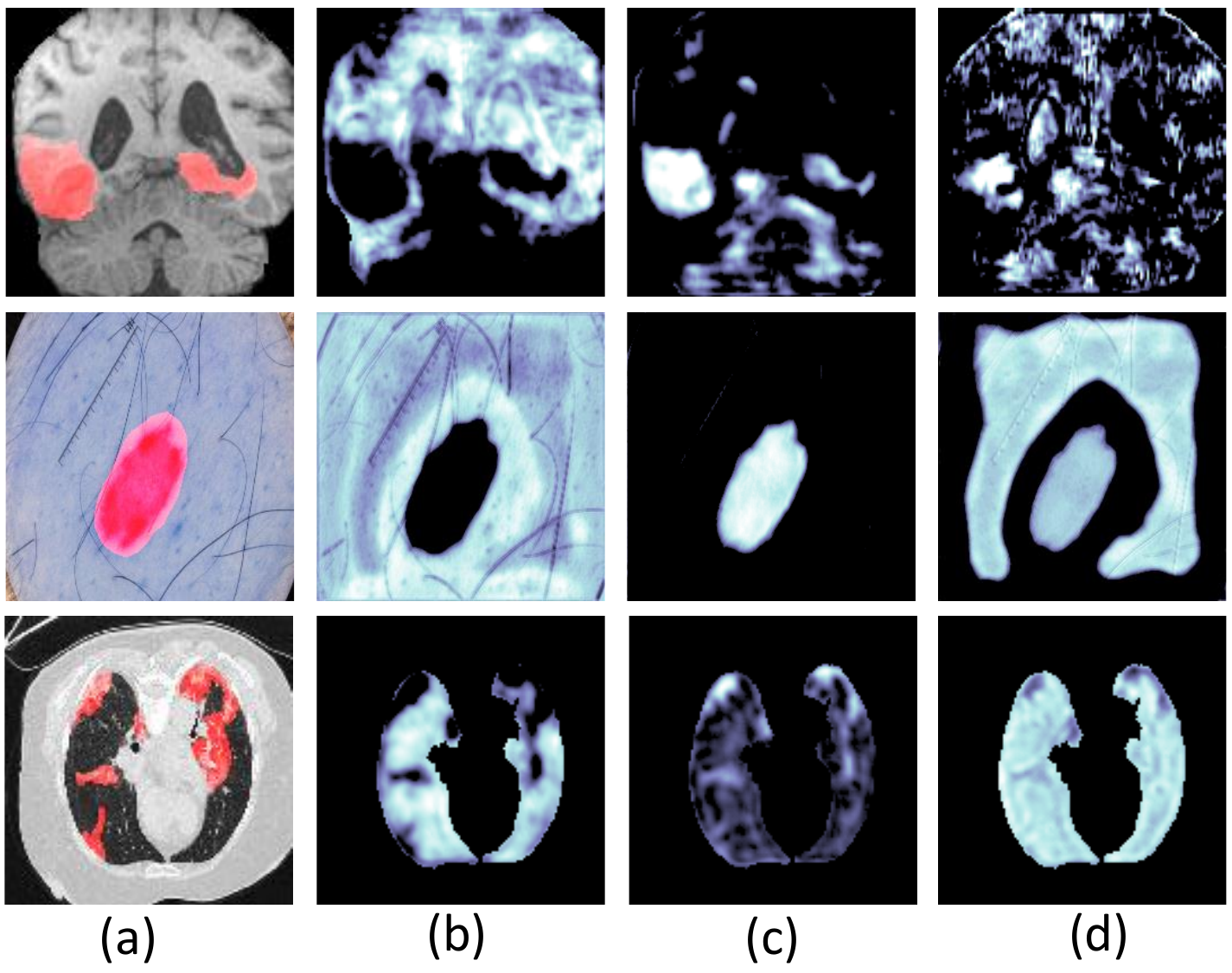}
    \caption{Visualization of the input images with object masks (a) and three corresponding deep features (b)-(d) extracted from three convolutional units. (b) shows that the activated values are on the background pixels/voxels, (c) shows the activated values are on the object pixels/voxels and (d) shows the activated values are on both object and background pixels/voxels.}
    \label{fig:featureresponse}
\end{figure}

However, it is not straightforward to apply these methods for interpreting the deep features learned for image segmentation.
The reasons are two-fold.
\begin{itemize}
    \item First, \textit{image recognition is a image-to-class problem while image segmentation is a image-to-image problem}.
    For image recognition, a neural network makes decisions based on the whole input image.
The size of the activation map for each class is the same as the size of the input image 
because almost all pixels/voxels are involved in the class recognition to some extent.
Thus, the binarized feature map~\citep{zhou2018interpreting} is to highlight some semantic regions which are meaningful for the corresponding recognized class.
However, image segmentation is a pixel/voxel-level classification problem which requires the neural network to recognize the label on each pixel/voxel. 
Therefore, the size of the activation map for each pixel/voxel is squeezed to that pixel/voxel.
From this point of view, the class activation map for each pixel/voxel is also a pixel/voxel, instead of a feature map.
    \item Second, \textit{all pixels/voxels of the input image are assumed to have one label for image recognition while they have different labels (object/background) for image segmentation}.
    For image classification, all pixels are assumed to have the same label thus the values computed by the CAM~\cite{zhou2016learning} or Grad-CAM~\cite{selvaraju2017grad} in the feature map indicate which pixels are more important for the final decision. 
   In other words, the values within one feature map are comparable since they work together to make one single decision.
However, for image segmentation, the network makes the decision based on each pixel and the values of the feature map do not reflect the importance of each pixel since different pixels carry different labels.
Thus, the separation ability of the deep feature is more important than the absolute values.
Overall, the absolute values of the deep features is meaningful for image recognition while the separation ability of the deep features is meaningful for image segmentation.
\end{itemize}

Fig.~\ref{fig:featureresponse}(b)-(d) give examples of three deep feature maps extracted on different units of convolutional layers on the trained neural networks from the input images (as shown in Fig.~\ref{fig:featureresponse}(a) with the ground-truth object masks shown in red).
We show that features with the high activated value on the background pixels/voxels (Fig.~\ref{fig:featureresponse}(b)), on the object pixels/voxels (Fig.~\ref{fig:featureresponse}(c)) and on both the object and background pixel/voxels (Fig.~\ref{fig:featureresponse}(d)).
The activated values (as measured by the CAM~\citep{zhou2016learning}) for objects are only high in Fig.~\ref{fig:featureresponse}(c).
However, the aim of segmentation is to separate the object and background pixels/voxels in the input images.
Thus, the feature maps on Fig.~\ref{fig:featureresponse}(b) are also discriminative since the object regions (with small responses) can be separated from the background regions (with large responses).
In fact, a negative weight can be applied on feature maps on Fig.~\ref{fig:featureresponse}(b) to highlight the object regions in the subsequent layers.
Therefore, the activated values in feature maps do not necessarily translate into the ability for segmentation but the segmentation ability of deep features is important for the final segmentation.

Another observation from Fig.~\ref{fig:featureresponse} is that the segmentation abilities of the deep features from different units are different.
For example, the segmentation abilities of the feature map shown on Fig.~\ref{fig:featureresponse}(b) and Fig.~\ref{fig:featureresponse}(c) are greater than the segmentation abilities of the feature map shown on Fig.~\ref{fig:featureresponse}(d).
In general, the segmentation abilities of deep features on different layers or on different units from the same layer are different and quantifying the segmentation abilities is important to understand the deep features and further understand the whole neural network for medical image segmentation.

\begin{figure}[!t]
    \centering
    \includegraphics[width=0.5\textwidth]{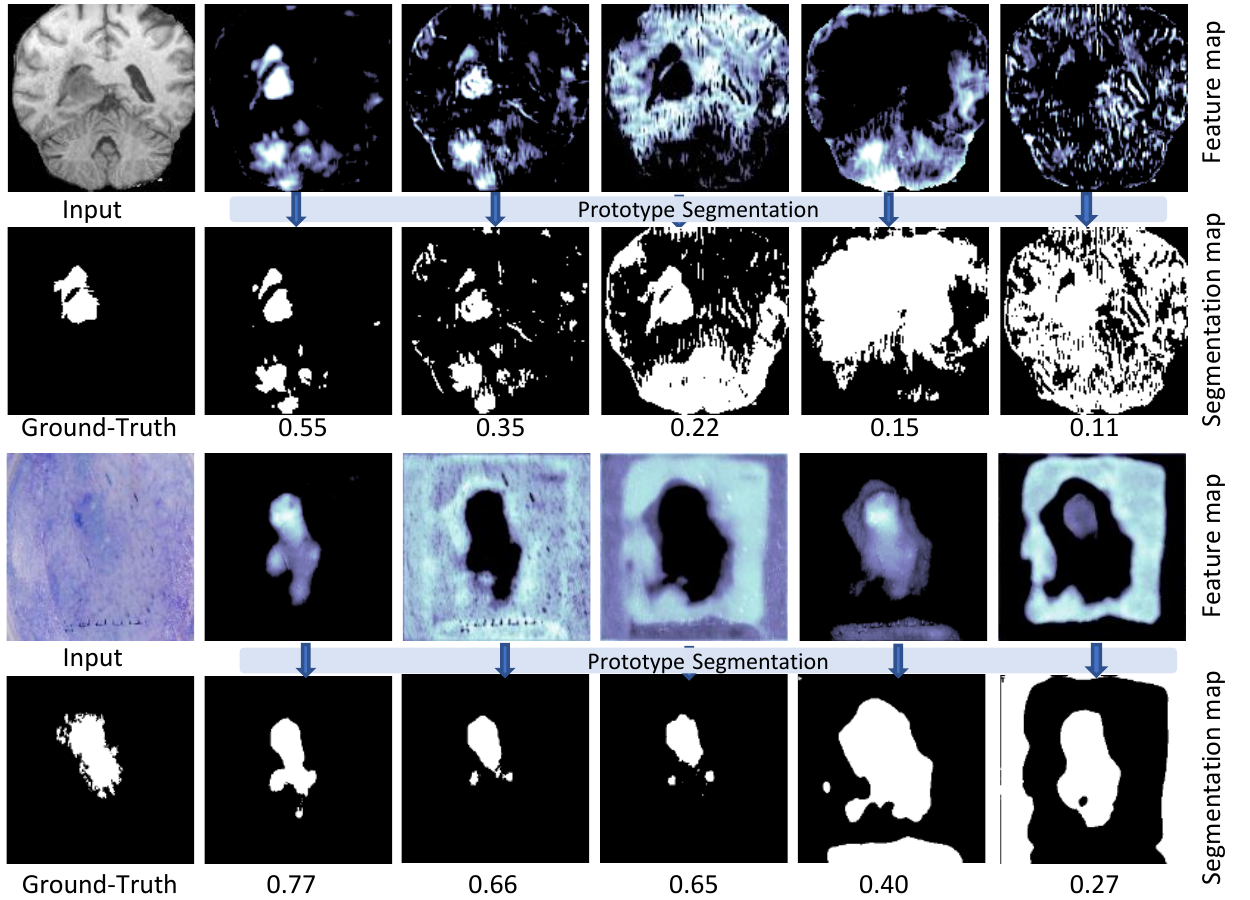}
    \caption{Examples of feature maps and their corresponding segmentation maps from the prototype segmentation for brain tumor (top two rows) and skin lesion (bottom two rows). The SA scores are shown on the bottom of each feature map.}
    \label{fig:examplefeature}
\end{figure}

\begin{figure*}[!t]
    \centering
    \includegraphics[width=\textwidth]{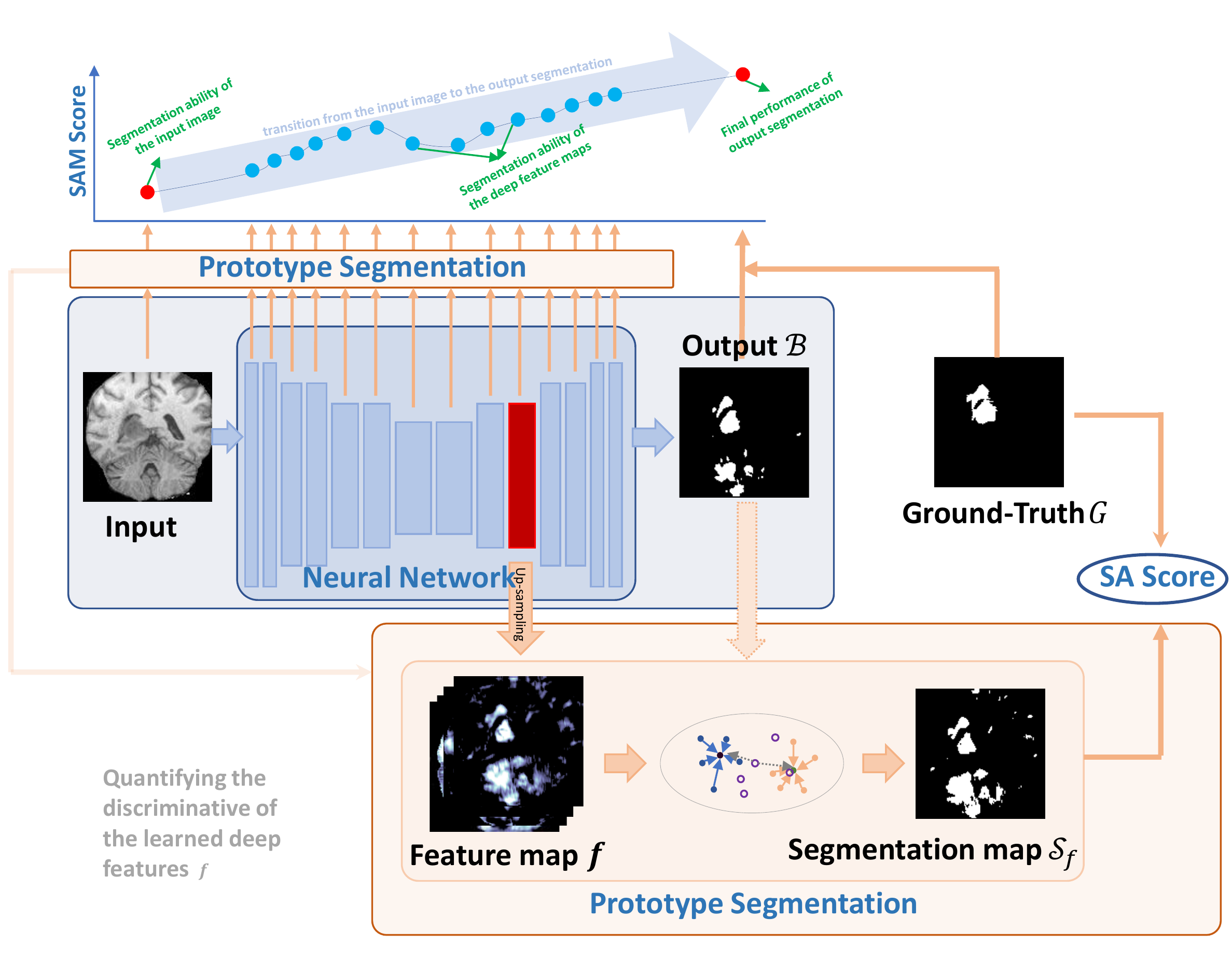}
    \caption{The framework of the proposed method for deep feature interpretation. Given any deep feature $f$ on the neural network, it is up-sampled to the size of the input image. Then the resized feature map is binarized by using the proposed prototype segmentation (ProtoSeg) with the initial mask $\mathcal{B}$. 
    The segmentation ability of the feature $f$ is measured by the SA score between the segmentation map $\mathcal{S}_f$ and the ground-truth $G$.
    Our proposed ProtoSeg can be used in any deep feature or input image to measure their segmentation abilities and reveal the transition of the segmentation ability from the input image to the output segmentation.}
    \label{fig:prototypescore}
\end{figure*}

\section{ProtoSeg: Prototype Segmentation on deep features}
In this section, we describe the proposed plug-and-play and parameter-free prototype segmentation (ProtoSeg) method which can compute a binary segmentation map on a given deep feature extracted from the neural network.

A deep neural network can be divided into two parts: a feature extractor $\mathbf{E_f}$ with parameters $\theta_f$ and a segmentation predictor $\mathbf{E_s}$ with parameters $\theta_s$.
Given the input image $x$, a deep feature map $f$ can be computed by: $f=\mathbf{E_f}(x,\theta_f)$ and the binarized segmentation output $\mathcal{B}$ can be obtained by $\mathcal{B}=\mathbf{E_s}(f,\theta_s)$.
Our aim is to quantify the segmentation ability of the deep feature $f$ by learning the $\mathbf{E_f}$ and $\mathbf{E_s}$ simultaneously, such that the segmentation predictor $\mathbf{E_s}$ can predict the label accurately.

\subsection{Parameter-free prototype segmentation}
We denote a learned deep feature tensor $f$ with the size of $H\times W\times C$ ($H$, $W$ and $C$ are the height, width and number of channels, respectively).
As we mentioned above,
a discriminative feature representation should have a small distance among pixels/voxels with the same label and should have a large distance among pixels/voxels with different labels.
This can be described as:
\begin{equation}
\label{eq:pairwise}
    \underset{\forall f_i, f_j, L(f_i)=L(f_j)}{\mathcal{D}(f_i,f_j)} < \underset{\forall f_i, f_j, L(f_i)\neq L(f_j)}{\mathcal{D}(f_i,f_j)}
\end{equation}
where $f_i$ and $f_j$ are the features at pixels/voxels $i$ and $j$ respectively.
$L(f_i)$ and $L(f_j)$ are the labels of the feature $f_i$ and $f_j$.
$\mathcal{D}$ is the distance function, such as the Euclidean distance used in this paper.
Ideally, the distance of any two pixels/voxels can be computed.
However, directly computing the distance from any given features $f_i$ and $f_j$ is inefficient since the size of feature maps is usually high which requires a large memory and a long computing time. 

In this paper, we resort to the prototype learning method~\citep{snell2017prototypical} to compute the distance among different locations in the deep features.
If the prototype (the center of the feature) for object and background regions are obtained, the Eq.~(\ref{eq:pairwise}) can be written as:
\begin{equation}
\label{eq:prototype}
    \underset{\forall f_i, \textbf{c}_k, L(f_i)=L(\textbf{c}_k)}{\mathcal{D}(f_i,\textbf{c}_k)} < \underset{\forall f_i, \textbf{c}_k, L(f_i)\neq L(\textbf{c}_k)}{\mathcal{D}(f_i,\textbf{c}_k)}
\end{equation}
where $\textbf{c}_k,k\in \{t,b\}$ are the prototypes of the object ($k=t$) or background regions which contain normal pixels/voxels ($k=b$). 
Eq.(\ref{eq:prototype}) shows that the distance of the features should be close to the center with the same label while far way from the center with different labels. 
For example, if $f_i$ is from the object region, 
we have:
\begin{equation}
    \mathcal{D}(f_i,\textbf{c}_t) < \mathcal{D}(f_i,\textbf{c}_b)
\end{equation}
which can be normalized by dividing the sum of these two distances:
\begin{equation}
    \frac{\mathcal{D}(f_i,\textbf{c}_t)}{\mathcal{D}(f_i,\textbf{c}_t)+\mathcal{D}(f_i,\textbf{c}_b)} < \frac{\mathcal{D}(f_i,\textbf{c}_b)}{\mathcal{D}(f_i,\textbf{c}_t)+\mathcal{D}(f_i,\textbf{c}_b)}
\end{equation}
When using the softmax function on both side, we can obtain:
\begin{equation}
\label{eq:softmax}
    \frac{\text{exp}\Big(-\mathcal{D}(f_i,\textbf{c}_t)\Big)}{\text{exp}\Big(-\big(\mathcal{D}(f_i,\textbf{c}_t)+\mathcal{D}(f_i,\textbf{c}_b)\big)\Big)} > \frac{\text{exp}\Big(-\mathcal{D}(f_i,\textbf{c}_b)\Big)}{\text{exp}\Big(-\big(\mathcal{D}(f_i,\textbf{c}_t)+\mathcal{D}(f_i,\textbf{c}_b)\big)\Big)}
\end{equation}
The Eq.(\ref{eq:softmax}) can be described as
$p(f_i,\mathbf{c}_t) > p(f_i,\mathbf{c}_b)$
for object feature $f_i$. 
Similarly, we can obtain $p(f_i,\mathbf{c}_t) < p(f_i,\mathbf{c}_b)$ for background feature $f_i$.

The prototype of the object $\mathbf{c}_t$ and background regions $\mathbf{c}_b$ can be computed by any cluster method, such as $k$-means.
However, using cluster methods is not efficient since the number of pixels/voxels is usually large. 
In this paper, we use the output of the neural network as the initial mask $\mathcal{B}\in[0,1]$ to compute the prototype of the object and background regions by:
\begin{equation}
    \begin{array}{cl}
        \mathbf{c}_t =& \sum (B_i * f_i)/\sum B_i  \\
        \mathbf{c}_b =& \sum \Big((1-B_i) * f_i\Big)/\sum (1-B_i)   \\
    \end{array}
\end{equation}
where $B_i\in \mathcal{B}$ and $B_i=1$ denotes as the object region while $B_i=0$ denotes the background region.
It is straightforward to generalize the proposed method into multi-label segmentation by adding new prototypes.

Once the prototype of the target and background is given, each pixel/voxel on the $i$th position can be classified based on a nearest neighbor over distance to the prototypes~\citep{snell2017prototypical}.
If $p(f_i,\mathbf{c}_t)>p(f_i,\mathbf{c}_b)$, the pixel/voxel belongs to the object.
Otherwise, it is from the background region.
Finally, a segmentation map $\mathcal{S}_{f}$ can be obtained based on the feature map $f$ by:
\begin{equation}
s_i=\left\{
    \begin{array}{ll}
        1, & \text{if} \ \ p_i(\mathbf{c}_t)>p_i(\mathbf{c}_b) \\
        0, & \text{otherwise} \\
    \end{array}
    \right.
\end{equation}
where $s_i\in \mathcal{S}_f$ is the segmentation on pixel/voxel $i$.

The advantage of the proposed prototype segmentation is that it is a parameter-free method and is differentiable and efficient to compute.
It aims to separate object and background regions in the feature maps on each input image, without considering their absolute value of responses.
As shown in Algorithm~\ref{alg:code}, our proposed ProtoSeg method is a plug-and-play module, which can be used in any deep feature or input image.

\begin{algorithm}[t]
\caption{ProtoSeg Pseudocode, PyTorch-like: \\
\url{https://github.com/shengfly/ProtoSeg}}
\label{alg:code}
\definecolor{codeblue}{rgb}{0.25,0.5,0.5}
\definecolor{codekw}{rgb}{0.85, 0.18, 0.50}
\lstset{
  backgroundcolor=\color{white},
  basicstyle=\fontsize{7.5pt}{7.5pt}\ttfamily\selectfont,
  columns=fullflexible,
  breaklines=true,
  captionpos=b,
  commentstyle=\fontsize{7.5pt}{7.5pt}\color{codeblue},
  keywordstyle=\fontsize{7.5pt}{7.5pt}\color{codekw},
}
\begin{lstlisting}[language=python]
# f: deep feature map
# b: initial mask in [0,1]. (0: background, 1: object)

def ProtoSeg(f,b): # compute the segmentation map
    c1 = Prototype(f,b) # prototype of object
    c2 = Prototype(f,1-b) # prototype of background
    
    p1 = -torch.pow(f-c1,2).sum()
    p2 = -torch.pow(f-c2,2).sum()
    
    m = torch.softmax([p2,p1],1) # the probability map
    
    #SAM = torch.argmax(m,1) # to compute the SAM
    
    return m

def Prototype(f,p):  # compute the prototype
    center = torch.sum(f*p)/torch.sum(p)
    return center
\end{lstlisting}
\end{algorithm}

\subsection{Network interpreting}

For any given deep feature $f$, the segmentation map $\mathcal{S}_{f}$ can be obtained by the proposed ProtoSeg method (Algorithm~\ref{alg:code}).
The segmentation ability of the deep feature $f$ can be measured by any metric used for segmentation evaluation.
In this paper, we use the well-established Dice score~\citep{milletari2016v} to compute the SA score for measuring the segmentation ability:
\begin{equation}
\label{eq:dice}
    \text{SA Score}(\mathcal{S}_f,G) = \frac{2|\mathcal{S}_f\cap G| }{|\mathcal{S}_f|+|G|}
\end{equation}
where $\mathcal{S}_f$ is the SAM obtained from the deep feature $f$ and $G$ is the ground-truth.
A high SA score indicates that the deep feature $f$ is discriminative and its corresponding SAM $\mathcal{S}_f$ is close to the ground-truth.
Therefore, $\text{SA Score}(\mathcal{S}_f,G)$ can be used to evaluate the segmentation ability of the feature map in a CNN, which is an objective score for interpretability that is comparable across layers and networks~\citep{zhou2018interpreting}.
Fig.~\ref{fig:examplefeature} shows several examples of feature maps $f$ and their corresponding SAM $\mathcal{S}_f$ computed by the proposed method.
The features can be sorted by the computed SA score.

Fig.~\ref{fig:prototypescore} shows the whole framework for network interpretation using the proposed plug-and-play ProtoSeg method, which can compute the segmentation ability on any given deep feature on the neural network.
 Our proposed ProtoSeg method can be divided into two methods: offline ProtoSeg and online ProtoSeg.
\begin{itemize}
    \item \textbf{Offline ProtoSeg}: The ProtoSeg method is applied to the trained neural network to measure the segmentation ability of deep features.
    \item \textbf{Online ProtoSeg}: The ProtoSeg method is differentiable and the SA Score can be used as a loss as part of the training with the neural network.
    Adding the SA Score in the training loss can increase the segmentation ability of the deep feature.
    
\end{itemize}

Using the online ProtoSeg can produce the segmentation maps of the deep features which are similar to the output/ground-truth. Thus, the segmentation maps of the learned deep features are more interpretable based on their similarity with the ground-truth. For example, if the SA score is low, the corresponding deep feature has a low segmentation ability which is reflected by the ProtoSeg loss in the training.

\section{Experiments of interpreting deep features}
In this section, we conduct experiments to use the proposed method to interpret the intermediate representation (deep features) learned by the deep neural networks for medical image segmentation.

\subsection{Datasets}

We use five datasets to evaluate the proposed method. 
In some datasets, the objects to be segmented are lesions and the background are normal regions. 
In other datasets, the objects are normal organs of interest while the background contains other structures in the image.
(1) BraTS~\citep{menze2014multimodal,bakas2017advancing} is used for brain tumor segmentation.
We use the subjects in BraTS18 as a training set and the new subjects in BraTS19 and BraTS20 as the testing set where subjects in the training set are excluded.
In total, 285 subjects are used for training and 84 subjects are used for testing.
Following the work~\citep{zhou2019unetplusplus}, we extract 2D slices from these patients with four modalities: FLAIR, T1, T1 with contrast (T1c) and T2 images, which are concatenated as the input.
We train networks to segment whole tumors, which considers all tumor labels as the positive class~\citep{zhou2019unetplusplus};
(2) ISIC~\citep{codella2018skin} is used for skin lesion segmentation.
We use the 2017 ISIC challenge dataset, which contains 2,000 dermoscopic images for training and 600 images for testing;
(3) COVID is used for COVID-19 infection segmentation on CT images.
We merge two public datasets: COVID-19 CT segmentation dataset~\footnote{\url{http://medicalsegmentation.com/covid19/}} which contains 100 CT slices and COVID-19 CT scans~\citep{ma2020towards} which contains 20 scans.
We collect 2D slices and split them into training (1,616 slices) and testing (328 slices) sets;
(4) Prostate~\citep{litjens2012pattern} is used for whole prostate segmentation on transverse T2-weighted scans and apparent diffusion coefficient (ADC) map.
We extract 2D slice from 32 scans and split them into training (375 slices) and testing (100 slices);
(5) Pancreas~\citep{dawant2007semi} is used for pancreatic parenchyma and pancreatic mass segmentation. 
We select 281 CT scans~\citep{simpson2019large} and randomly split them into training (6,860 slices) and test (1,692 slices) sets.


\subsection{Network training}
The Adam optimizer in the PyTorch package is used to train the U-Net neural network, with a batch size of 10.
The learning rate is set to 0.0001 and reduced to half at every 20 epochs.
Networks for all the datasets are trained with a total of 50 epochs with the widely used cross-entropy loss.
To evaluate the interpretation difference between offline ProtoSeg and online ProtoSeg,
we train the model without and with the loss of the SA Score.
For training without the SA Score, the training loss is applied only on the output image: $\mathcal{L}=\mathcal{L}_g$.
For training with the SA Score loss, we also apply the SA Score on the feature segmentation map:
$\mathcal{L}=\mathcal{L}_g+\sum \text{SA Score}/N$ where $N$ is the number of feature segmentation maps.
For a fair comparison, we use the same parameter and training methods for neural networks on all five datasets.
No additional re-sampling or post-processing is included.~\citep{isensee2018nnu}.

\subsection{Interpreting deep features of U-Net}
\begin{figure}[!t]
    \centering
    \includegraphics[width=0.5\textwidth]{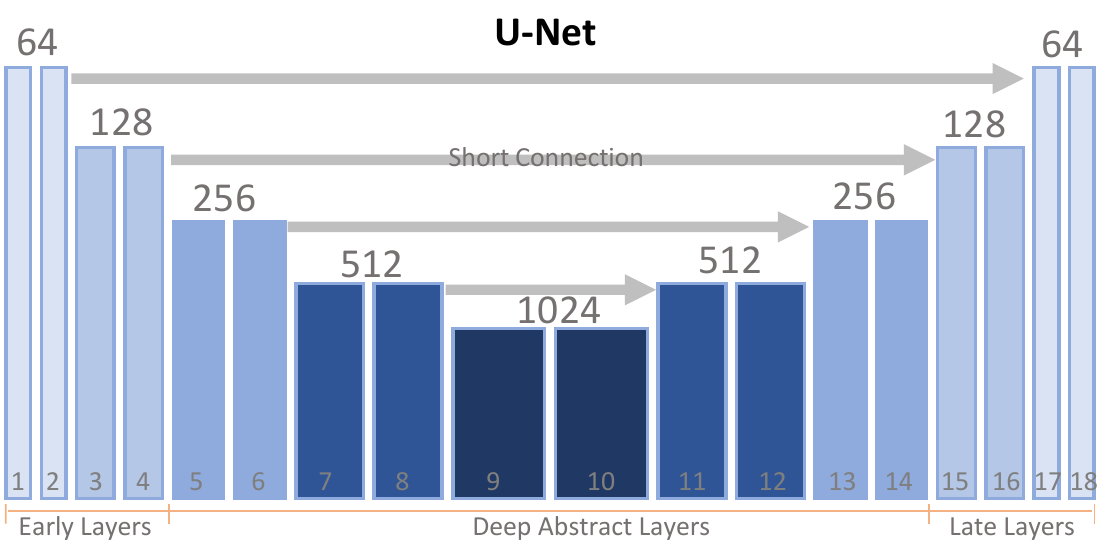}
    \caption{The structure of the U-Net. Typically, there are 18 convolutional layers, indexed by the number from 1 to 18. The number of channels on each layer is indicated on the top. We divide the features of each layer into three groups: early features (from 1-4 layers), deep abstract features (from 5-14 layers) and the late features (the last 4 layers).}
    \label{fig:UnetStructure}
\end{figure}

Although our method can be used in any neural network, we focus on the widely used segmentation neural network U-Net~\citep{ronneberger2015u} in this paper.
As shown in Fig.~\ref{fig:UnetStructure}, a typical U-Net contains 18 feature maps, denoted by $f_i$, $i=1,2,...,18$.
The spatial resolutions of these features are different due to the max-pooling and up-sampling operations.
The numbers of channels of these feature maps are also various (shown in Fig.~\ref{fig:UnetStructure}).
As with U-Net~\citep{ronneberger2015u}, all convolutional layers (except the last one) have the same parameters: kernel size is 3$\times$3, stride is 1 and padding is 1.
The kernel size of 1$\times$1 is used on the last output layer for predicting the output which is used as an initial mask in the proposed prototype segmentation.
Batch normalization~\citep{ioffe2015batch} is used to accelerate the training and the default activation function is the widely used rectified linear unit (ReLU).
As shown in Fig.~\ref{fig:UnetStructure}, we divide the 18 deep layers into three groups: early layers which contain the first four convolutional layers, deep abstract layers which contain the layers from 5-14 layers and late layers which contain the last four layers.
The feature maps on the early and late layers have a high spatial resolution but a small number of feature dimensions.
On the other hand, feature maps on the deep abstract layers have a low spatial resolution but a large number of channels.

\begin{figure*}[!t]
    \centering
    \includegraphics[width=\textwidth]{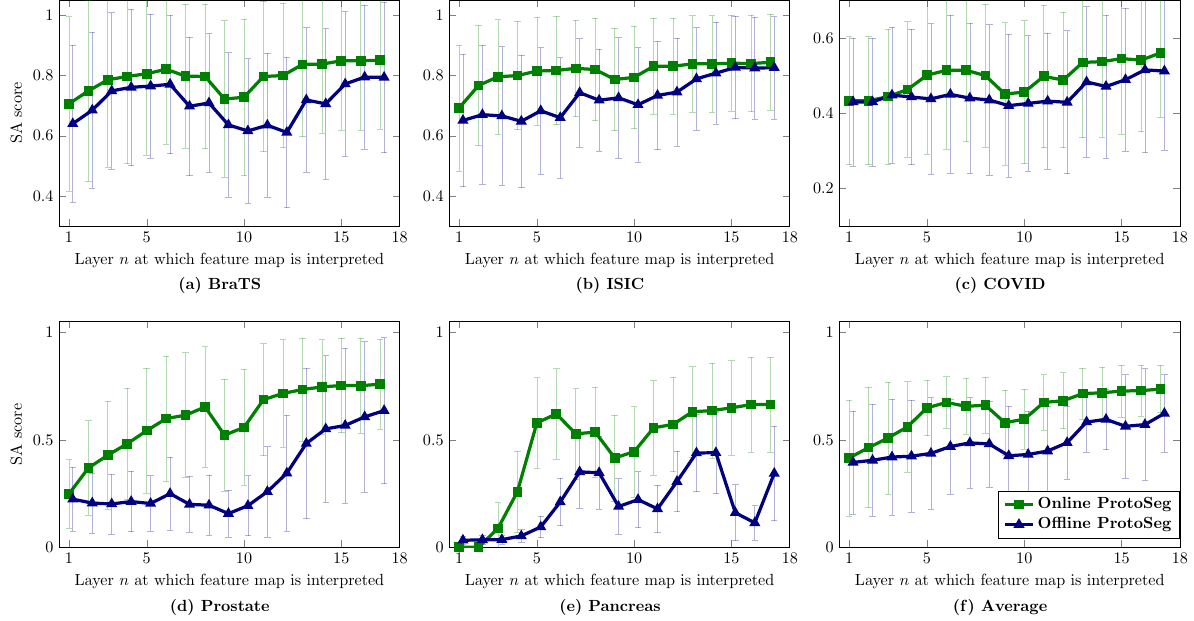}
    \caption{The segmentation ability (SA score) of deep features from different layers of the U-Net. Training the network with the SA score in the loss can increase the segmentation ability of deep features.}
    \label{fig:unetlayers}
\end{figure*}

\begin{figure*}[!t]
    \centering
    \includegraphics[width=\textwidth]{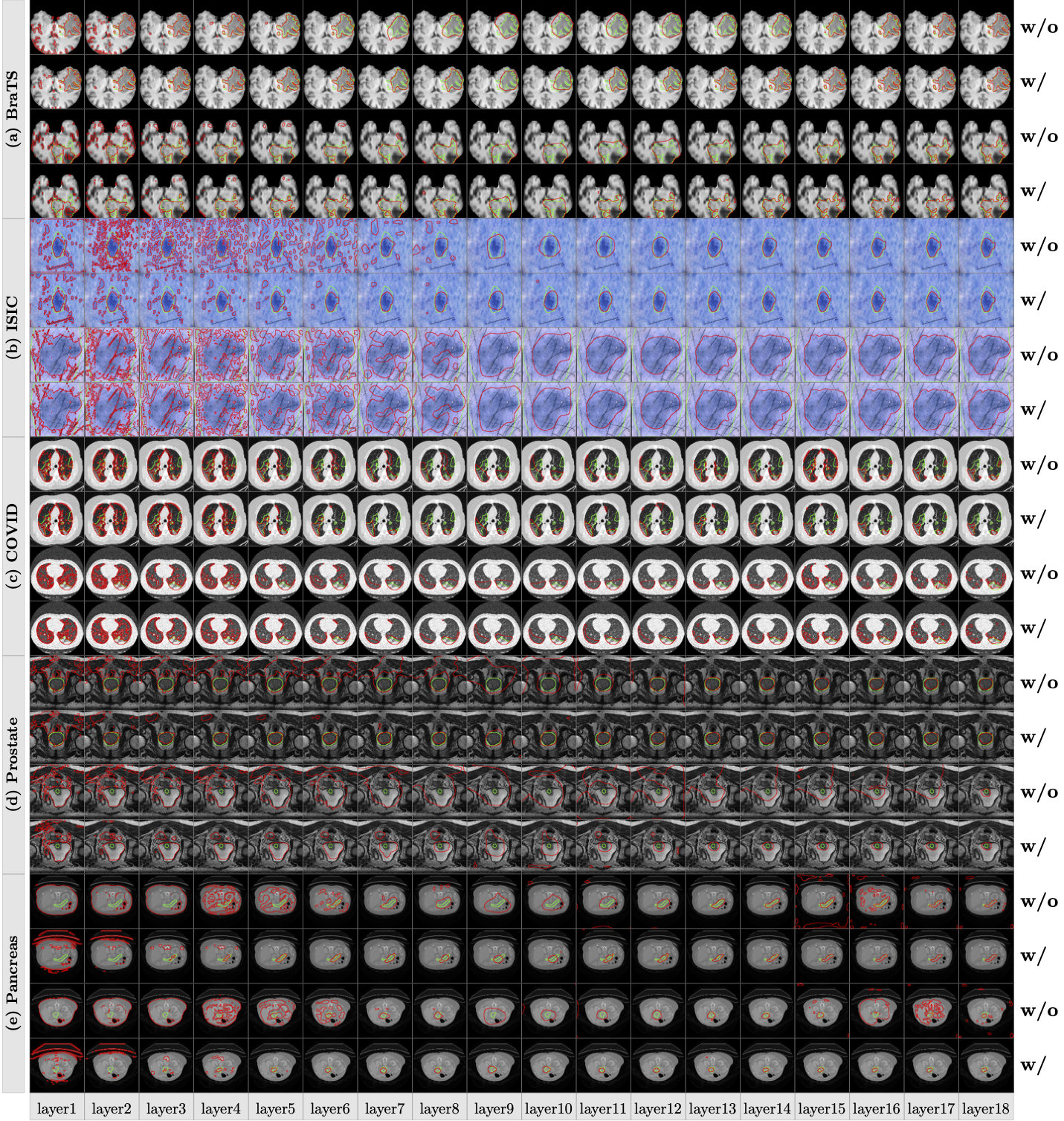}
    \caption{Visualizations of the segmentation maps obtained from different feature maps on layers 1-18 of the U-net. The green and red contours denote the ground-truth and the results of the segmentation maps. w/ means the network is trained with the SA score in the training loss while w/o means without the SA score.}
    \label{fig:layerimageseg}
\end{figure*}

\begin{table*}[!t]
    \centering
    \caption{The performance of different methods on five datasets. Output $\mathcal{B}$: the output of the U-Net. Segmentation $\mathcal{S}_{f_{18}}$: the SAM on the last layer (the 18-th layer in U-Net). It shows that training the network with the SA score in the loss does not hurt the performance.}
    \label{tab:DiceUnetLayer}
    \begin{tabular}{c|cc|cc}
    \toprule
    \multirow{2}{*}{Dataset} & \multicolumn{2}{c|}{Offline ProtoSeg} & \multicolumn{2}{c}{Online ProtoSeg} \\
    \cmidrule{2-5}
    & Output $\mathcal{B}$ & Segmentation $\mathcal{S}_{f_{18}}$ &Output $\mathcal{B}$ & Segmentation $\mathcal{S}_{f_{18}}$ \\
    \midrule
    BraTS & 0.8471$\pm$0.23 & 0.7944$\pm$0.25 & 0.8428$\pm$0.24 & \textbf{0.8514}$\pm$0.23\\
    ISIC & 0.8330$\pm$0.18 & 0.8260$\pm$0.17 & 0.8353$\pm$0.17 & \textbf{0.8445}$\pm$0.16\\
    COVID & 0.4862$\pm$0.22 & 0.5123$\pm$0.21 & 0.4840$\pm$0.22 & \textbf{0.5597}$\pm$0.17\\
    Prostate & 0.7232$\pm$0.28 & 0.6365$\pm$0.34 & 0.7557$\pm$0.20 & \textbf{0.7585}$\pm$0.21\\
    Pancreas & 0.6475$\pm$0.25 & 0.3439$\pm$0.22 & 0.6441$\pm$0.23 & \textbf{0.6635}$\pm$0.22 \\
    \bottomrule
    \end{tabular}
\end{table*}

Deep feature maps are obtained after each layer of the U-Net.
We use the proposed ProtoSeg method to compute the SAMs on these deep feature maps and then compute their corresponding SA scores referring to the ground-truth as the measurements of the segmentation ability.

\subsection{Feature segmentation ability of different layers in U-Net}

\subsubsection{Our method can reveal the transition of the segmentation ability from early to late layers}

Our proposed method can evaluate the feature segmentation ability on any given layer.
In this section, we compute the segmentation ability of all the 18 layers on U-Net.
For deep features on each layer, the SAM is computed by the proposed ProtoSeg and the corresponding SA score is calculated based on the ground-truth. 
Fig.~\ref{fig:unetlayers} shows the average SA scores on the testing set of SAMs from different layers.
The figure shows that the SA score increases from early to late layers, except on the several layers on deep abstract layers which have a low spatial resolution, yielding a small SA score.
Our results confirm that the segmentation ability of deep features learned for segmentation shows a transition from early to late layers in the U-Net.

\subsubsection{Our method can visualize the SAMs of deep features}

Our proposed method can also visualize the SAMs of deep features from different layers and
Fig.~\ref{fig:layerimageseg} shows several examples from different datasets.
On early layers, deep features contain detailed information of texture, boundary and spatial structure, and they are sensitive to the noise on the input images and these noises are gradually eliminated when layers on the neural network go deep.
The SA score is low on deep abstract layers and the main reason is that the spatial resolution is low.
However, the low spatial deep features after max-pooling can localize the object regions, representing the semantic context for segmentation.
The boundaries on these deep abstract layers are smooth around the object regions and the detailed boundaries are recovered on the following late layers. 
When combining the features of the early and deep abstract layers, the segmentation maps from the late layers are close to the ground-truth.

\subsubsection{Training the network with the SA score (online ProtoSeg) in the loss does not hurt the performance}

Table~\ref{tab:DiceUnetLayer} shows the SA scores of neural network's output $\mathcal{B}$ and the SA scores on the feature segmentation maps from the last layer $\mathcal{S}_{f_{18}}$ (the 18-th layer in Fig.~\ref{fig:UnetStructure}).
We choose the last layer because it provides the highest SA score (as shown in Fig.~\ref{fig:unetlayers}).
There are two observations: 
(1) using the SA score in the training loss on the SAMs does not hurt the accuracy of the neural network.
On the five datasets, there is no significant difference between the outputs $\mathcal{B}$ with/without using the SA score in the training loss on deep features.
However, (2) the segmentation on feature maps $\mathcal{S}_{f_{18}}$ using the SA score in the training loss provides a higher SA score than the output of the neural network, which demonstrates that the deep features contain rich information and the proposed ProtoSeg could find the deep features with the highest SA score.
Fig.~\ref{fig:unetlayers} and \ref{fig:layerimageseg} show that even without the ProtoSeg loss, the intermediate features computed by different convolutional kernels/filters have some correlations with the final output. The reason would be that U-Net decomposes the input lesion-bearing image, and Fig.~\ref{fig:layerimageseg} shows that the abstract layers help reduce the noise and locate the main part of the lesions or target regions.
Layer 17 picks high-intensity regions, because these regions had a high segmentation ability as measured by the proposed SA score, not because these regions had high intensities. 

\subsubsection{Training the network with the SA score (online ProtoSeg) in the loss can increase the segmentation ability of deep features}

Fig.~\ref{fig:layerimageseg} and Table~\ref{tab:DiceUnetLayer} show that the segmentation ability of the deep features trained with the SA scores (online ProtoSeg) in the loss are higher than the ones trained without SA score (offline ProtoSeg), indicating that using the SA score in the loss can increase the segmentation ability of deep features on the intermediate layers of the U-Net.
 In addition, the SA scores of the segmentation $\mathcal{S}_{18}$ are higher than the segmentation of the output $\mathcal{B}$ on these five datasets.

\subsubsection{Discussion}

Several insights can be found on Fig.~\ref{fig:unetlayers}, Fig.~\ref{fig:layerimageseg} and Table~\ref{tab:DiceUnetLayer} for understanding the U-Net.
There is a transition of the segmentation ability from the early layers to the late layers.
When reducing the spatial resolution, the down-sample layers can gradually eliminate the noise and roughly localize the object regions.
The detailed boundaries of the object shape are gradually recovered in the late layers.
It is the first time to quantify and visualize the widely-suspected phenomenon that the encoder path in U-Net is to encode the input images into the global context to gradually eliminate the noise and the decoder path is to recover the object shape from the global context and the early feature maps.
\textit{In other words, downsampling is important for eliminating the noise or false positive.}
In addition, the main reason why applying the ProtoSeg on the last feature map can provide better performance than the output of the neural network is that the prototypes of the object and background regions are learned directly from the feature maps which are discriminative  and adaptive for each input image.
However, the output $\mathcal{B}$ is computed by the linear weights which are learned from the whole dataset thus less adaptive to each image.

\subsection{Feature segmentation ability of different units on the last layer}

\begin{figure}[!t]
    \centering
    \includegraphics[width=0.5\textwidth]{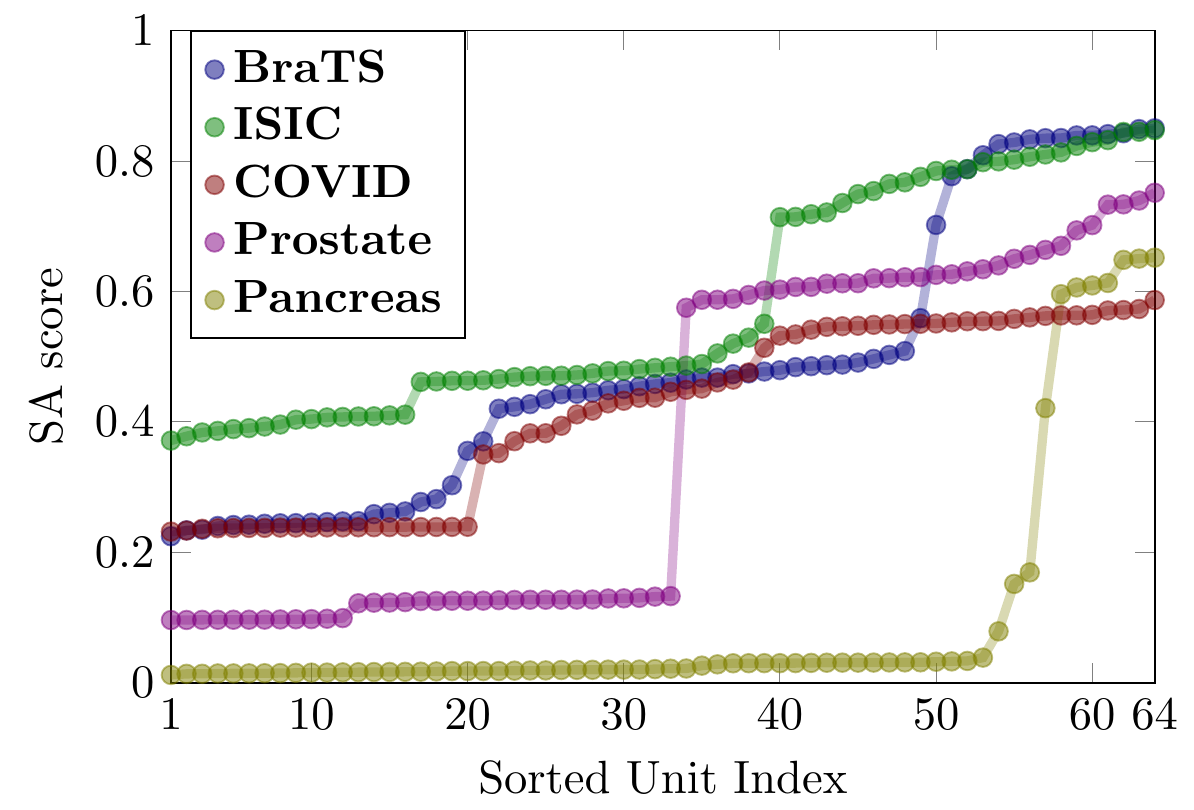}
    \caption{The sorted SA score of 64 units on the last convolutional layer of the U-Net. }
    \label{fig:unitscores}
\end{figure}

\begin{figure}[!ht]
    \centering
    \includegraphics[width=0.5\textwidth]{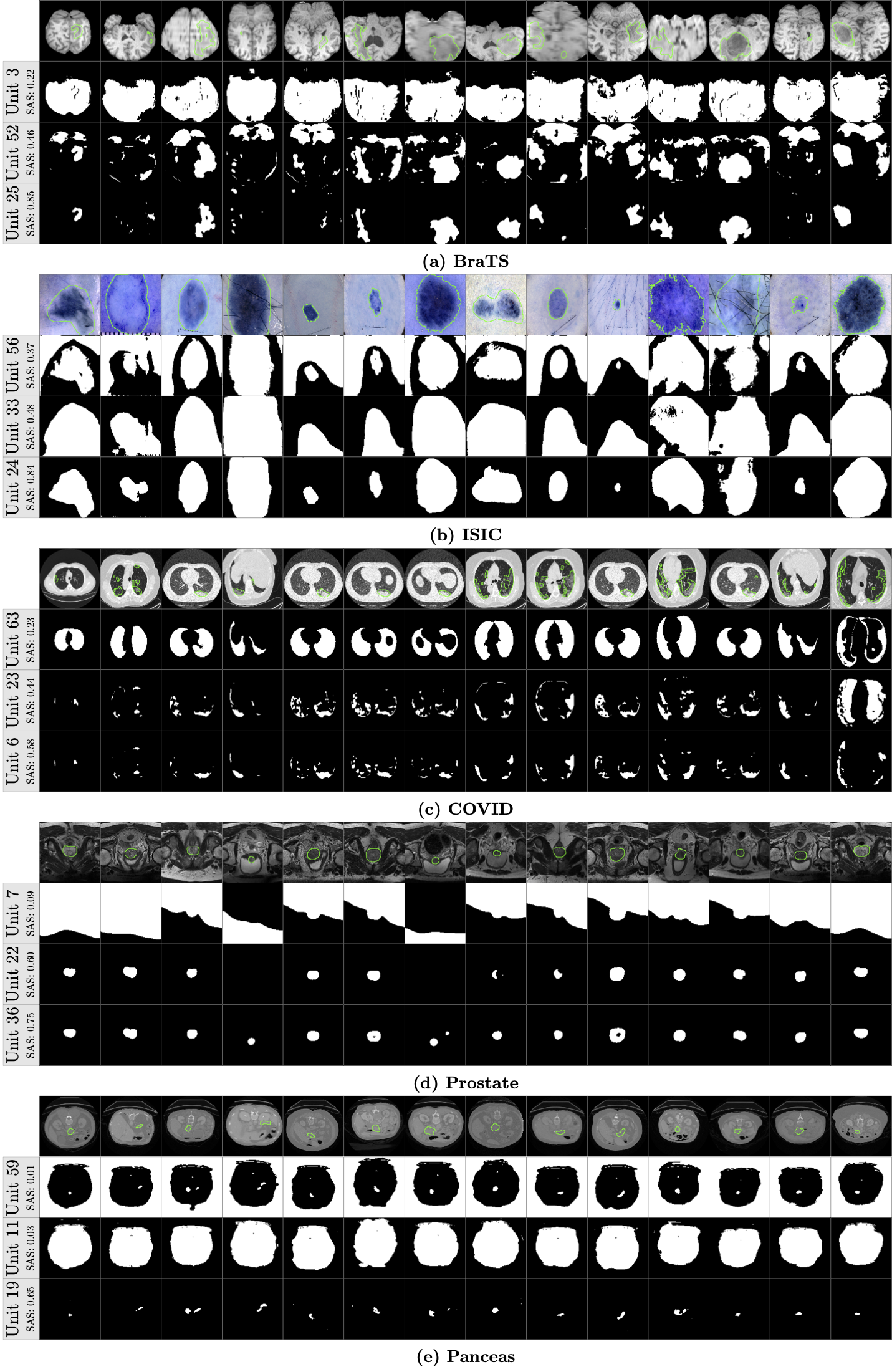}
    \caption{Several examples of the segmentation maps from different units. The green contours denote the ground-truth and the SAS is the SA score of each unit.}
    \label{fig:unitsegmap}
\end{figure}

\begin{figure}[!t]
    \centering
    \includegraphics[width=0.5\textwidth]{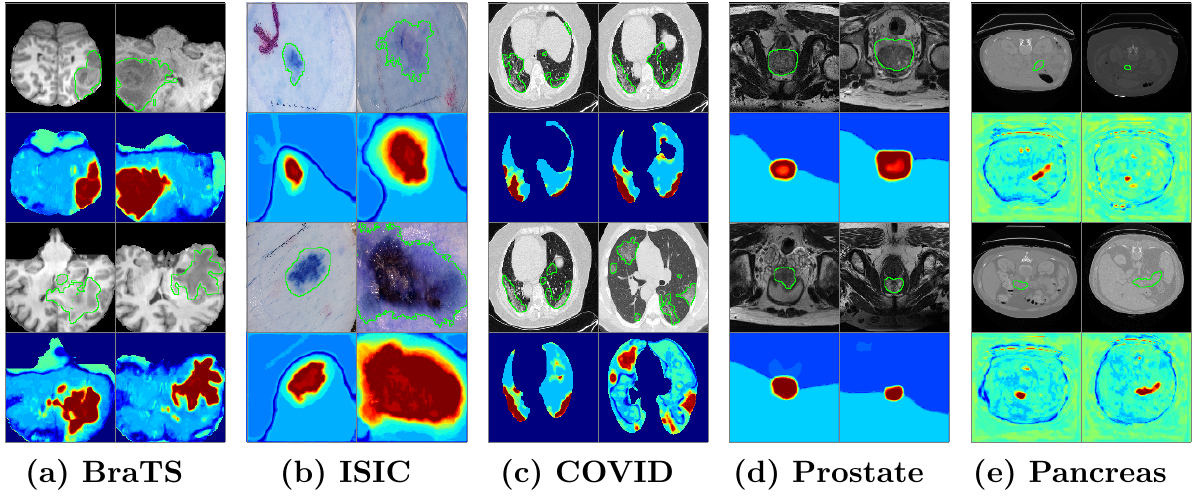}
    \caption{The heat-map of the average of unit SAMs on the last layer of U-Net.
    The first and third row show the input images with ground-truth (green contour), and the second and the fourth row show the corresponding heat-map.}
    \label{fig:averageunit}
\end{figure}

Although our proposed method can evaluate the segmentation ability of units at any layer,
we focus on the last layer because the SA score of the last layer is usually higher than other layers (as shown in Fig.~\ref{fig:unetlayers}).
There are 64 units on the last convolutional layer (layer 18 in Fig.~\ref{fig:UnetStructure}) and it is interesting to know the segmentation ability of these individual units.
We extract the deep feature on each unit and use the proposed ProtoSeg to compute the SAM and the SA score.
Fig.~\ref{fig:unitscores} shows the SA scores of 64 different units on the last layer.
The index of the units is meaningless since the weights are randomly initialized.
Thus, we sort the units according to their SA scores.
It clearly shows that units can be grouped into two parts according to the SA scores and only 20\%-50\% units have higher SA scores close to the performance of the output.
Fig.~\ref{fig:unitsegmap} shows the segmentation maps of several selected units with the lowest, mediate and highest SA scores.
Fig.~\ref{fig:averageunit} shows the average response (heat-map) of the segmentation maps computed on the units of the last layer.
Object pixels/voxels have high values, indicating that most units can separate object regions from the background regions.
There is a clear transition from the object regions to the background regions and the pixels/voxels on the boundary of object regions have higher uncertainty values~\citep{nair2020exploring}.

Our results support the theory that the last layer of the U-Net is heavily oversized and only a part of units have higher discriminative scores~\citep{hu2016network}.
From Fig.~\ref{fig:unitsegmap} we can see that the segmentation maps of the weak units with low SA scores contain both the object and background regions.
The SA score of the best unit is similar to the output of the neural network which linearly combines all 64 units on the last layer.

One possible reason is that 
the network tends to overfit on a group of units with high SA scores and their SAMs are close to the ground-truth on most images.
The output of the neural network can be denoted as:
\begin{equation}
    y = \sum_i^{N}w_i \cdot f_i + b
\end{equation}
where $w_i$ is the learned weight of the unit, $b$ is the bias and $f_i$ is the deep feature.
The units can be divided into active and inertia groups:
\begin{equation}
\label{eq:lazy}
    y = \underbrace{\sum_i^{N_1}w_i \cdot f_i}_\text{active}+\underbrace{\sum_{j=N_1}^{N}w_j \cdot f_j}_\text{inertia}+b
\end{equation}
During training,  the active units provide similar results to the ground-truth which will dominate the gradient toward the weights of the active units.
Meanwhile, the inertia units give a constant output $\sum_{j=N_1}^{N}w_j \cdot f_j \approx C$ (with low discriminative SA score) for both the object and background pixels/voxels.
Eq.~(\ref{eq:lazy}) becomes:
\begin{equation}
    \begin{array}{cc}
        y \approx & \underbrace{\sum_i^{N_1}w_i \cdot f_i}_\text{active}+\underbrace{C}_\text{inertia}+b \\
        \approx & \underbrace{\sum_i^{N_1}w_i \cdot f_i}_\text{active} + b_1
    \end{array}
\end{equation}
where $b_1$ is the bias: $b_1=C+b$ and $N_1 < N$ is the number of active units.
From Fig.~\ref{fig:unitscores} we can see that the number of active units varies by datasets.
For example, on the ISIC dataset, there are more than 20 active units while on the Pancreas dataset in which images contain small objects there are only around 11 active units.
Our results demonstrate that network pruning might be helpful for building efficient segmentation networks which can exclude the inertia units by network pruning method~\citep{molchanov2016pruning,zhu2017prune}.

\begin{figure*}[!t]
    \centering
    \includegraphics[width=\textwidth]{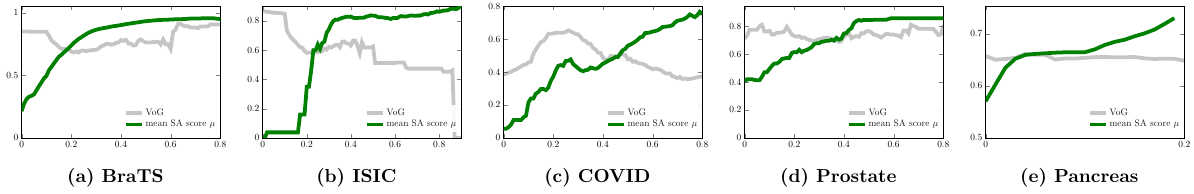}
    \caption{The relationship between the Dice accuracy and scores computed by the proposed method (the  mean SA score $\mu$, the green lines) and the VoG score (the gray lines) across subjects in each dataset. The $y$ axis is the Dice accuracy of the neural network's output $\mathcal{B}$  and the $x$ axis is the  mean SA score $\mu$ and the scaled VoG score for improving the visualization. }
    \label{fig:allconsistant}
\end{figure*}

\begin{figure*}[!t]
    \centering
    \includegraphics[width=\textwidth]{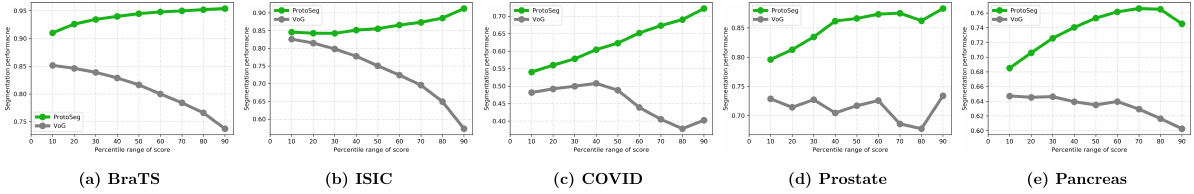}
    \caption{ The segmentation performance (y-axis) for the test images without ground-truth thresholded by the percentile of the mean SA score $\mu$ (green lines) and the VoG score (gray lines) (x-axis). Across the five datasets, the segmentation accuracy increases with an increase in the mean SA scores but not the VoG scores.}  
    \label{fig:plotpercentage}
\end{figure*}

\begin{figure}[!t]
    \centering
    \includegraphics[width=0.5\textwidth]{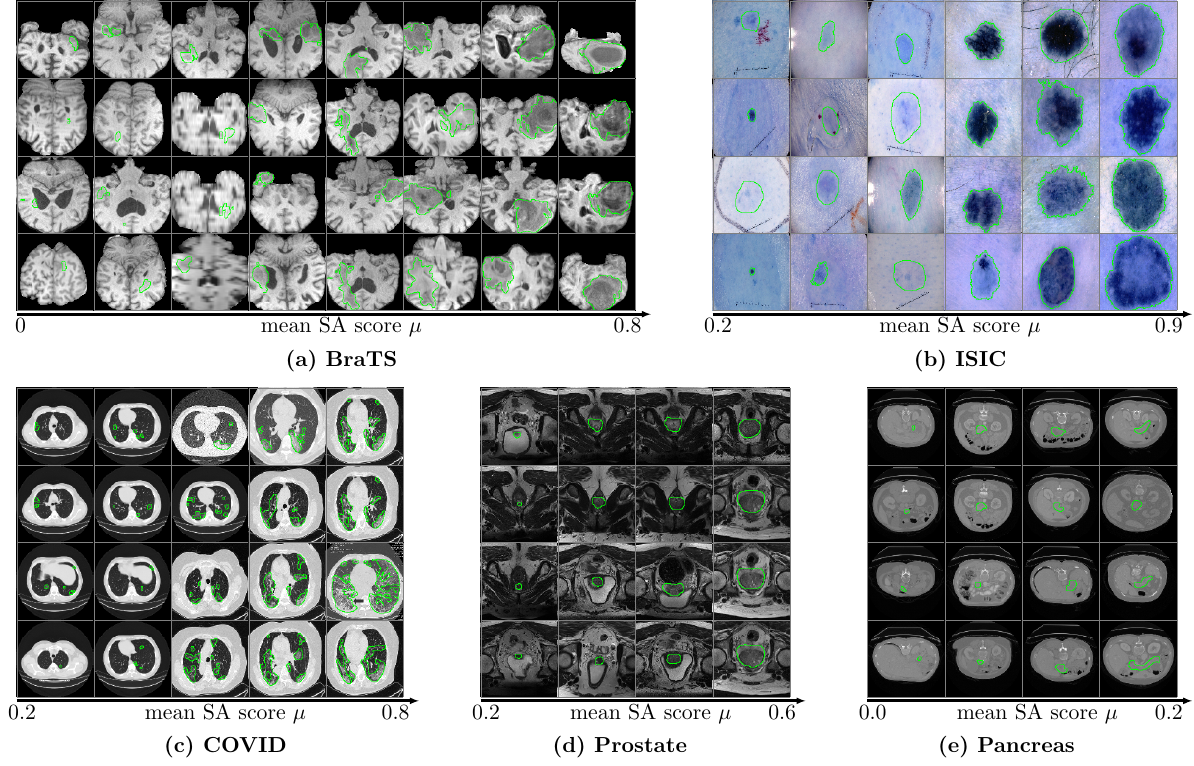}
    \caption{ The ranking list of images based on the mean SA score $\mu$ on the five datasets. The segmentation maps of the images with a low mean SA score $\mu$ need more attention.}
    \label{fig:imageconsistency}
\end{figure}

\begin{figure*}[!t]
    \centering
    \includegraphics[width=\textwidth]{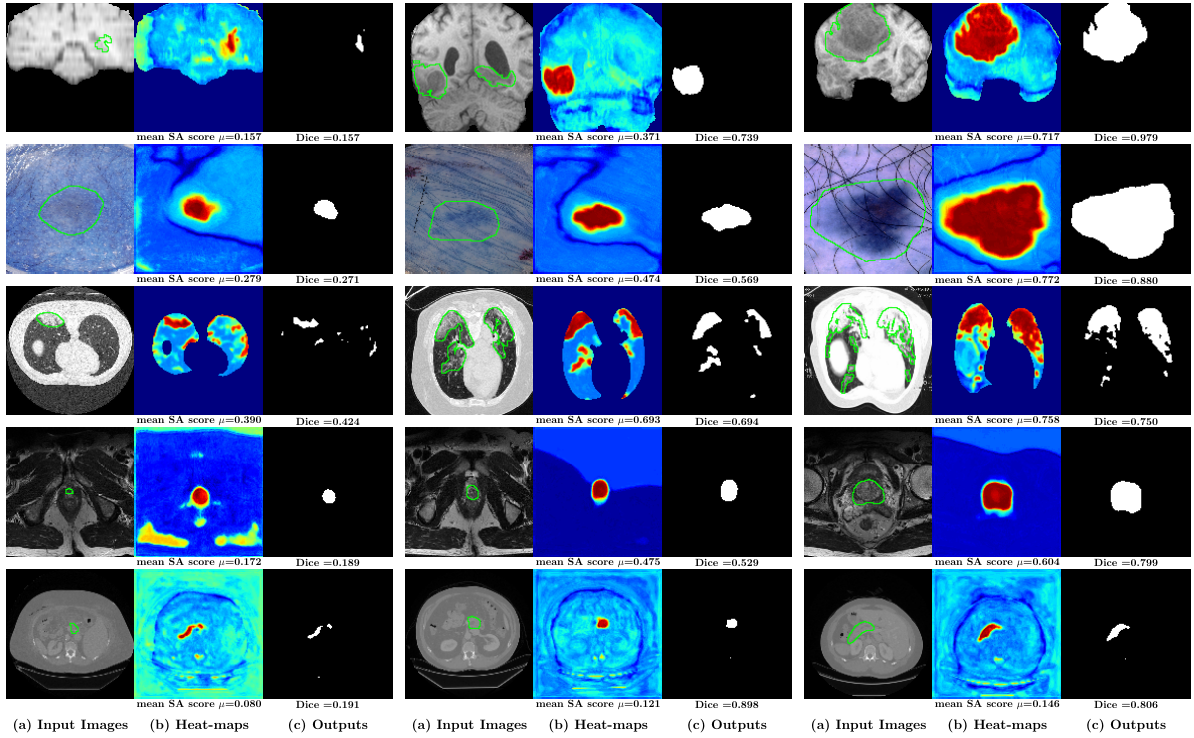}
    \caption{Examples of the (a) input images (with ground-truth highlighted with green contours), the corresponding (b) heat-maps computed based on the segmentation of units on last two years and (c) outputs of neural network. The corresponding mean SA score and Dice score are shown on the bottom of each image.}
    \label{fig:visualizedconsistency}
\end{figure*}

\begin{figure}[!ht]
    \centering
    \includegraphics[width=0.5\textwidth]{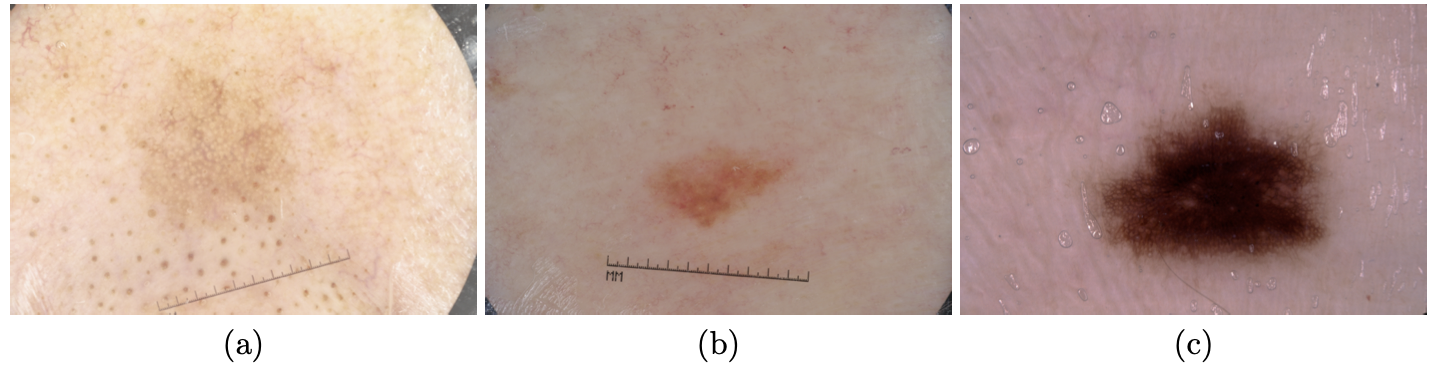}
    \caption{Examples of different input images on the ISIC dataset. The separableness is: (a) $<$ (b) $<$ (c). }
    \label{fig:separablenss}
\end{figure}

\begin{figure*}[!h]
    \centering
    \includegraphics[width=\textwidth]{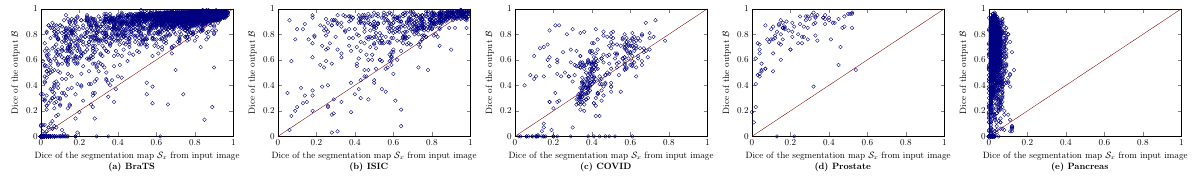}
    \caption{The scatter plot of the Dice score of the neural network versus the Dice score of segmentation map on the input image on different datasets.}
    \label{fig:inputimageseg}
\end{figure*}
\subsection{Evaluate the quality of segmentation outputs}
Most studies for medical image segmentation only report the average Dice score for system evaluation.
However, the performance can vary on different input images.
For example, the performance is high on some images with high resolution and clear boundary of the objects while the Dice score is low on some images with poor quality.
Given an image from the testing set without the ground-truth, it is important to know the estimated accuracy of its segmentation output from the neural network.
In other words, a confident score of each test image is important for end-users to make decisions with a ``human-in-the-loop" workflow.

In practice, there is no ground-truth to compute the confident score of the predicted segmentation map in the test set.
In this section, we define a measurement for roughly evaluating the quality of the output segmentation $\mathcal{B}(x)$, based on the SA score between the feature segmentation $\mathcal{S}_f(x)$ maps and neural network output $\mathcal{B}(x)$ given the input image $x$.
We compute the SA score of the feature segmentation on each unit of the last two layers, considering the output $\mathcal{B}(x)$ as the best approximation to the unknown ground-truth.
The mean SA score $\mu(x)\in[0,1]$ is defined as:
\begin{equation}
    \mu(x) = \frac{1}{N}\sum_i^N \text{SA score}\Big(\mathcal{S}_{f_i}(x),\mathcal{B}(x)\Big)
\end{equation}
where $N$ is the number of units on the last two layers involved in the computation and $f_i$ is the deep feature on the $i$th unit.

Fig.~\ref{fig:allconsistant} shows the relationship between the performance of the neural network and the  mean SA score $\mu$ of the units from the last two convolutional layers.
\textit{The segmentation performance of the network correlates to the  mean SA score $\mu$ on all five datasets}. 
For images with a high  mean SA score $\mu$, the Dice score of the neural network's output $\mathcal{B}$ is also high.
We also compare it with the Variance of Gradients (VoG)~\citep{agarwal2022estimating} which is used for ranking the test images. VoG is specifically designed for image classification. Fig.10 shows that the correlations between the VoG score and segmentation accuracy (Dice) are smaller than the correlations between the mean SA score $\mu$ and segmentation accuracy for medical image segmentation on the five datasets.
Fig.~\ref{fig:imageconsistency} shows the ranking list of images based on the mean SA score $\mu$.
Images with a low mean SA score $\mu$ is the difficult images which need more attention in practice.
Thus, the mean SA score is a meaningful measurement for ranking the test images without ground-truth by difficulty and surfacing a tractable subset of the most difficult test images for human-in-the-loop auditing~\citep{agarwal2022estimating}.
Fig.~\ref{fig:visualizedconsistency} shows examples of the image with heat-maps and outputs from the neural network.
The corresponding mean SA scores of segmentation maps on deep features and Dice scores of the neural network's output are also shown on each image.
From the example images we can see that images with small objects or smooth boundaries usually have a low  mean SA score $\mu$ and the segmentation performance of the output is also low.
Therefore, the performance of segmentation is mainly affected by the small objects or objects with the smooth boundaries which usually have a small  mean SA score $\mu$.
Put simply, smaller and blurry objects are more difficult to segment.

Based on the above observation, the  mean SA score $\mu$ can be used for sample rejection, which can configure the system to warn the end-users for the prediction on any images with the mean SA score lower than a threshold. The neural network usually has a low performance on images with small lesions or object regions with smooth boundaries and our proposed mean SA score $\mu$ can be served as a rough measurement for the end-users to pay more attention to these images.
Similar to the work~\citep{ghesu2020quantifying}, we also use the term \textbf{coverage}, as an expected percentile of samples to be rejected for further attention on these segmentation results.
For example, at a coverage of 100\%, the neural network outputs its prediction on all test images, while at coverage of 80\%, the neural network rejects to output the segmentation on 20\% of images with the lowest  mean SA score $\mu$~\citep{ghesu2020quantifying}.
By rejecting the images with small objects or unclear boundaries indicated by the  mean SA score $\mu$, the performance of the U-Net increases on the remaining test images, as shown in Table~\ref{tab:coverageper}.
 In Fig.~\ref{fig:plotpercentage}, we plot the segmentation accuracy of examples bucketed by the mean SA score $\mu$, inspired by~\cite{agarwal2022estimating}. 
It shows that samples with lowest percentiles of mean SA score $\mu$ have low segmentation accuracy and the segmentation performance increases with an increase in mean SA score $\mu$, which are consistent across the five datasets.
 In addition, Fig.~\ref{fig:plotpercentage} shows that the VoG score is not a good predictor for the segmentation accuracy and segmentation performance decreases with an increase in VoG score.

\begin{table*}[!t]
    \centering
    \caption{The performance of segmentation with sample coverage rates of 100\%,90\%, 75\% and 50\%.}
    \label{tab:coverageper}
    \begin{tabular}{c|cccc}
    \toprule
    Coverage &  100\% & 90\% & 70\% & 50\% \\
    \midrule
    BraTS &  0.8428$\pm$0.24 & 0.9102$\pm$0.11 & 0.9344$\pm$0.06 & 0.9446$\pm$0.04\\
    ISIC & 0.8353$\pm$0.17 & 0.8457$\pm$0.16 & 0.8421$\pm$0.16 & 0.8551$\pm$0.15\\
    COVID &  0.4880$\pm$0.22  & 0.5400$\pm$0.17 & 0.5778$\pm$0.15 & 0.6229$\pm$0.13 \\
    Prostate & 0.7557$\pm$0.20 & 0.7959$\pm$0.16 & 0.8344$\pm$0.14 & 0.8660$\pm$0.08 \\
    Pancreas & 0.6441$\pm$0.23 & 0.6850$\pm$0.19 & 0.7255$\pm$0.16 & 0.7529$\pm$0.15\\
    \bottomrule
    \end{tabular}
\end{table*}

\subsection{Understanding the separableness of the input images}
The separablenesses of the input images is defined as the segmentation ability of the input images by the color/intensity values directly.
For example, the image in Fig.~\ref{fig:separablenss}(a) has a lower separablenesses since the color values on the object regions are quite similar to the color values on the normal regions.
It is hard to segment such lesions, even by human experts.
However, the image on Fig.~\ref{fig:separablenss}(c) has the largest separablenesses because color values on the lesion regions (the object) are quite different from the color values on the normal regions (the background).

The separableness is considered as the special segmentation ability of the input image, which can also be computed by the proposed ProtoSeg method.
The segmentation map $\mathcal{S}_x$ can be computed by considering the colors/intensities of input image $x$ as features.
Thus, the performance of the segmentation map $\mathcal{S}_x$ indicates the quality or the separableness of the input image.
Lesions in some images have high contrast and clear boundary, which are easy to be segmented, even using traditional methods based on color or intensity values, yielding a SAM with a high SA score.

In this section, we study the relationship of performance between the $\mathcal{S}_x$ (the segmentation map computed directly on input image $x$) and the network's output $\mathcal{B}(x)$, investigating how neural networks improve the segmentation on each individual image $x$.
Fig.~\ref{fig:inputimageseg} shows the relationship between the segmentation map of input image $\mathcal{S}_x$ and the corresponding output of the neural network $\mathcal{B}(x)$.
In most cases, U-net can improve the segmentation performance and the Dice score of the output $\mathcal{B}(x)$ is higher than the SA score of $\mathcal{S}_x$.
We define the distance between the Dice scores of the $\mathcal{S}_x$ and $\mathcal{B}(x)$ on one image as: $d=\text{Dice}(\mathcal{B},G)-\text{SA score}(\mathcal{S}_x,G)$ where $G$ is the ground-truth.
For one dataset, we compute the average distance over all possible test images as: $m(d)=\sum d/N$.
A high $m(d)$ means a large gain obtained by training a deep network for segmentation.
Table~\ref{tab:insegdice} shows the Dice scores of the $\mathcal{S}_x$, $\mathcal{B}$ and the $m(d)$ on different datasets.
The most challenging dataset among the five datasets is the COVID segmentation, which has the lowest $m(d)$, indicating that the gain achieved by the trained U-Net is low.
In addition, we have found that most images on BraTS and ISIC are easy samples in which the lesion regions can be easily separated only based on the input image pixels, yielding higher SA scores of $\mathcal{S}_x$.
Thus, our proposed method can be used to describe the characteristics of the dataset for segmentation.

\begin{table}[!t]
    \centering
    \caption{The average of the distance between the Dice scores of network output $\mathcal{B}$ and segmentation map $\mathcal{S}_x$ based on colors/intensities of input image.}
    \label{tab:insegdice}
    \begin{tabular}{c|ccc}
    \toprule
    Dataset & $\mathcal{S}_x$ & $\mathcal{B}$ &  $m(d)$ \\
    \midrule
    BraTS &  0.657 & 0.842 & 0.185\\
    ISIC  &  0.667 & 0.835 & 0.168\\
    COVID &  0.411 & 0.484 & 0.072\\
    Prostate & 0.230 & 0.757 & 0.525\\
    Pancreas & 0.030 & 0.644 & 0.613\\
    \bottomrule
    \end{tabular}
\end{table}

\subsection{Understanding the feature segmentation ability with noise inputs}

\begin{figure*}[!t]
    \centering
    \includegraphics[width=\textwidth]{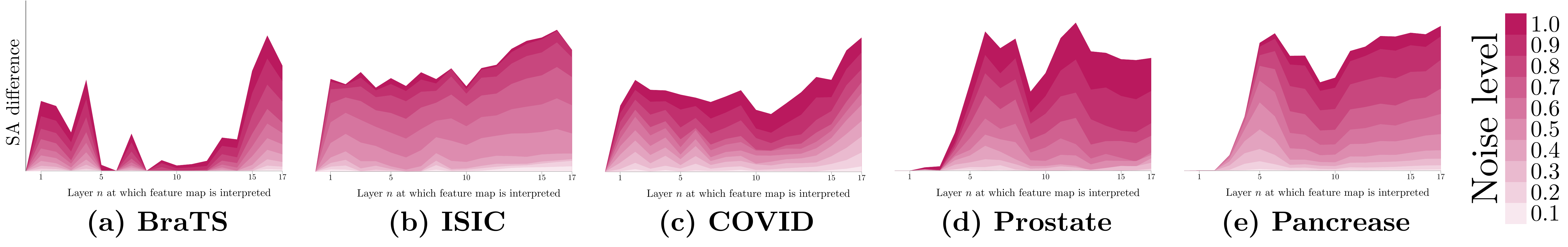}
    \caption{The segmentation ability (SA) differences on different layers of U-Net between the noise and the clean inputs. The different noise levels are indicated by different colors.}
    \label{fig:noise}
\end{figure*}

The segmentation ability (SA) score is also a useful measurement to investigate how the segmentation ability of feature changes when the input changes, such as with the different levels of noise.
This section presents the SA difference between the noise and clean input images.
For each input image, we add the random noise on each pixel/voxel with different levels (the maximum value of the noise) from 0.1 to 1. 
The SA difference is computed by: $SA(i_n)-SA(i)$ where $i_n$ is the noise version of the input image $i$.
Fig.~\ref{fig:noise} shows the results on the five datasets.
On BraTS, the noise affects the segmentation ability on the early and late layers of U-Net.
On the other four datasets, the noise affects the segmentation ability on all layers and the abstract layers are less sensitive to noise compared to early and late layers.
The results of this figure show that the effects of the noise are different when U-Net is trained on different datasets.
Another finding is that the late layers are more sensitive to the noise of the input images which is introduced by the shortcut connection for the early layers in U-Net.
Thus, the shortcut connection in U-Net needs to be further investigated by model developers.

\section{Conclusions and future work}
This paper presents a prototype segmentation (ProtoSeg) method which can compute the SAM on deep features extracted from neural networks.
Our proposed method can be used to evaluate the segmentation ability of any deep feature.
We have used it to evaluate the discriminative score of different layers of the most popular U-Net and the score of different units on the last layer of the U-Net.
We have found that there is a transition of the segmentation ability from the early layers to the late layers and the down-sampling on the U-Net is important to eliminate the noise and localize the lesions.
Fig.~\ref{fig:layerimageseg} shows that \textbf{the U-Net is actually a denoise model, which reduces the noise on the input image toward the final segmentation map}.
We have further proposed to define the  mean SA score $\mu$ between the output of the neural network and the feature segmentation maps to evaluate the quality of the input images. Most studies for segmentation only report the average Dice performance on the whole test images.  However, the accuracy of some images (e.g., those with small lesions or lesions with smooth boundaries) is low. Without the ground-truth on the test images, it is hard for end-users to know which segmentation results need to be double-checked.  
Our experimental results show that the mean SA score is linearly correlated to the accuracy of segmentation from the neural network, and therefore, SA can serve as a quantity to automatically reject some images that may be subject to large segmentation errors even when the ground-truth is not available. 
When the low SA values trigger the rejection of these images, end-users need to pay more attention to the segmentation results of these images.
In addition, we have used the proposed method to evaluate the separableness of each individual image based on segmentation map of colors/intensities in the input images and further used it to describe the characteristics of the dataset for medical image segmentation.

Experimental results show that using the ProtoSeg loss in the training does not hurt the performance but increases the segmentation ability of the deep features.
Since the ProtoSeg loss is not applied to the final output, it does not affect the accuracy of the output. Using the ProtoSeg loss can produce the segmentation maps of the learned deep features which are similar to the output/ground-truth. 
This provides a better interpretation by measuring the similarity between the segmentation maps and outputs. Thus, using the ProtoSeg loss can increase the interpretability of the deep features. 
On the other hand, a low SA score of the deep learned features indicate that it is hard to interpret those deep features since they are not similar to the output.
In addition, Fig.~\ref{fig:layerimageseg} and Fig.~\ref{fig:unitscores} produced by the proposed method show that the features from different layers or on the same layer are redundant in U-Net and neural network pruning is needed in future to build efficient models.

We have found that using the ProtoSeg loss does not improve the final accuracy, indicating that the intermediate losses on feature maps do not affect the performance of the segmentation. The possible reason is that adding the ProtoSeg loss only increase the interpretability of deep features which may not transfer to the accuracy of the output.

The limitations of the paper include:
(1) we only used the Dice score as the SA score to measure the segmentation ability of the deep feature since the Dice score is a widely used metric for training and evaluation.
Other metrics can be used, such as Jaccard index, pixelwise accuracy, sensitivity and specificity~\citep{li2020transformation}, which will be our future work.
(2) We only evaluated the proposed method on 2D network models. While we expect the conclusion to be general on 2D and 3D images, and for 2D and 3D network models, our future work includes testing this plug-and-play module into 3D networks.
(3) we did not find any relation between the SA scores and the corresponding learned weights on the units of the last layer since our proposed ProtoSeg focuses on the separable ability of the features instead of the absolute response on each unit.
Thus, a unit with a high SA score may have low response values and thus need a high weight to output the segmentation map.

The proposed ProtoSeg method aims to understand the learned deep features of the neural network, which can be considered as an auditing tool to provide a quantity value of the segmentation ability of the deep feature in different layers. It has the potential application to further quantitatively measure the sensitivity of input images with small variations and noise, dataset shift, out-of-sample effects, etc . 
For example, Fig.~\ref{fig:noise} shows how the segmentation ability of deep features in different layers changes with different level of noise.
Fig.~\ref{fig:inputimageseg} shows that the images on BraTS are easy to segment based on input intensities and Fig.~\ref{fig:noise} further verifies that it may not necessary to use a deep network since the abstract layers are less sensitive to the input changes.
Other potential applications of the proposed method include the automatical evaluation of the feature maps on any network or block for segmentation.
When model developers design new modules, the proposed method can be used to evaluate how the new module affects or boosts performance.
Thus it can provide insights for interpreting the network or segmentation results.
One of the future works is to extend the proposed ProtoSeg method to interpret the deep features of neural networks for classification which needs different prototype computations of the target class and the background.
Our study motivates future work towards building more explainable AI systems for medical image segmentation,
helping model developers to understand the transition of segmentation ability from input to output, and ranking the test images without ground-truth by the difficulty of segmentation and automatically suggesting the most challenging examples for  human experts to review.

\bibliographystyle{model2-names.bst}
\biboptions{authoryear}
\bibliography{IntUnet.bib}

\begin{thebibliography}{33}
\expandafter\ifx\csname natexlab\endcsname\relax\def\natexlab#1{#1}\fi
\providecommand{\url}[1]{\texttt{#1}}
\providecommand{\href}[2]{#2}
\providecommand{\path}[1]{#1}
\providecommand{\DOIprefix}{doi:}
\providecommand{\ArXivprefix}{arXiv:}
\providecommand{\URLprefix}{URL: }
\providecommand{\Pubmedprefix}{pmid:}
\providecommand{\doi}[1]{\href{http://dx.doi.org/#1}{\path{#1}}}
\providecommand{\Pubmed}[1]{\href{pmid:#1}{\path{#1}}}
\providecommand{\bibinfo}[2]{#2}
\ifx\xfnm\relax \def\xfnm[#1]{\unskip,\space#1}\fi
\bibitem[{Agarwal et~al.(2022)Agarwal, D'souza and
  Hooker}]{agarwal2022estimating}
\bibinfo{author}{Agarwal, C.}, \bibinfo{author}{D'souza, D.},
  \bibinfo{author}{Hooker, S.}, \bibinfo{year}{2022}.
\newblock \bibinfo{title}{Estimating example difficulty using variance of
  gradients}, in: \bibinfo{booktitle}{Proceedings of the IEEE/CVF Conference on
  Computer Vision and Pattern Recognition}, pp. \bibinfo{pages}{10368--10378}.
\bibitem[{Arrieta et~al.(2020)Arrieta, D{\'\i}az-Rodr{\'\i}guez, Del~Ser,
  Bennetot, Tabik, Barbado, Garc{\'\i}a, Gil-L{\'o}pez, Molina, Benjamins
  et~al.}]{arrieta2020explainable}
\bibinfo{author}{Arrieta, A.B.}, \bibinfo{author}{D{\'\i}az-Rodr{\'\i}guez,
  N.}, \bibinfo{author}{Del~Ser, J.}, \bibinfo{author}{Bennetot, A.},
  \bibinfo{author}{Tabik, S.}, \bibinfo{author}{Barbado, A.},
  \bibinfo{author}{Garc{\'\i}a, S.}, \bibinfo{author}{Gil-L{\'o}pez, S.},
  \bibinfo{author}{Molina, D.}, \bibinfo{author}{Benjamins, R.}, et~al.,
  \bibinfo{year}{2020}.
\newblock \bibinfo{title}{{Explainable Artificial Intelligence (XAI): Concepts,
  taxonomies, opportunities and challenges toward responsible AI}}.
\newblock \bibinfo{journal}{Information fusion} \bibinfo{volume}{58},
  \bibinfo{pages}{82--115}.
\bibitem[{Bakas et~al.(2017)Bakas, Akbari, Sotiras, Bilello, Rozycki, Kirby,
  Freymann, Farahani and Davatzikos}]{bakas2017advancing}
\bibinfo{author}{Bakas, S.}, \bibinfo{author}{Akbari, H.},
  \bibinfo{author}{Sotiras, A.}, \bibinfo{author}{Bilello, M.},
  \bibinfo{author}{Rozycki, M.}, \bibinfo{author}{Kirby, J.S.},
  \bibinfo{author}{Freymann, J.B.}, \bibinfo{author}{Farahani, K.},
  \bibinfo{author}{Davatzikos, C.}, \bibinfo{year}{2017}.
\newblock \bibinfo{title}{Advancing the cancer genome atlas glioma mri
  collections with expert segmentation labels and radiomic features}.
\newblock \bibinfo{journal}{Scientific data} \bibinfo{volume}{4},
  \bibinfo{pages}{170117}.
\bibitem[{Chen et~al.(2020)Chen, Bei and Rudin}]{chen2020concept}
\bibinfo{author}{Chen, Z.}, \bibinfo{author}{Bei, Y.}, \bibinfo{author}{Rudin,
  C.}, \bibinfo{year}{2020}.
\newblock \bibinfo{title}{Concept whitening for interpretable image
  recognition}.
\newblock \bibinfo{journal}{Nature Machine Intelligence} .
\bibitem[{Codella et~al.(2018)Codella, Gutman, Celebi, Helba, Marchetti, Dusza,
  Kalloo, Liopyris, Mishra, Kittler et~al.}]{codella2018skin}
\bibinfo{author}{Codella, N.C.}, \bibinfo{author}{Gutman, D.},
  \bibinfo{author}{Celebi, M.E.}, \bibinfo{author}{Helba, B.},
  \bibinfo{author}{Marchetti, M.A.}, \bibinfo{author}{Dusza, S.W.},
  \bibinfo{author}{Kalloo, A.}, \bibinfo{author}{Liopyris, K.},
  \bibinfo{author}{Mishra, N.}, \bibinfo{author}{Kittler, H.}, et~al.,
  \bibinfo{year}{2018}.
\newblock \bibinfo{title}{Skin lesion analysis toward melanoma detection: A
  challenge at the 2017 international symposium on biomedical imaging (isbi),
  hosted by the international skin imaging collaboration (isic)}, in:
  \bibinfo{booktitle}{International Symposium on Biomedical Imaging (ISBI
  2018)}, pp. \bibinfo{pages}{168--172}.
\bibitem[{Dawant et~al.(2007)Dawant, Li, Lennon and Li}]{dawant2007semi}
\bibinfo{author}{Dawant, B.M.}, \bibinfo{author}{Li, R.},
  \bibinfo{author}{Lennon, B.}, \bibinfo{author}{Li, S.}, \bibinfo{year}{2007}.
\newblock \bibinfo{title}{Semi-automatic segmentation of the liver and its
  evaluation on the {MICCAI} 2007 grand challenge data set}.
\newblock \bibinfo{journal}{3D Segmentation in The Clinic: A Grand Challenge} ,
  \bibinfo{pages}{215--221}.
\bibitem[{Ghesu et~al.(2020)Ghesu, Georgescu, Mansoor, Yoo, Gibson, Vishwanath,
  Balachandran, Balter, Cao, Singh et~al.}]{ghesu2020quantifying}
\bibinfo{author}{Ghesu, F.C.}, \bibinfo{author}{Georgescu, B.},
  \bibinfo{author}{Mansoor, A.}, \bibinfo{author}{Yoo, Y.},
  \bibinfo{author}{Gibson, E.}, \bibinfo{author}{Vishwanath, R.},
  \bibinfo{author}{Balachandran, A.}, \bibinfo{author}{Balter, J.M.},
  \bibinfo{author}{Cao, Y.}, \bibinfo{author}{Singh, R.}, et~al.,
  \bibinfo{year}{2020}.
\newblock \bibinfo{title}{Quantifying and leveraging predictive uncertainty for
  medical image assessment}.
\newblock \bibinfo{journal}{Medical Image Analysis} \bibinfo{volume}{68},
  \bibinfo{pages}{101855}.
\bibitem[{Gu et~al.(2020)Gu, Wang, Song, Huang, Aertsen, Deprest, Ourselin,
  Vercauteren and Zhang}]{gu2020net}
\bibinfo{author}{Gu, R.}, \bibinfo{author}{Wang, G.}, \bibinfo{author}{Song,
  T.}, \bibinfo{author}{Huang, R.}, \bibinfo{author}{Aertsen, M.},
  \bibinfo{author}{Deprest, J.}, \bibinfo{author}{Ourselin, S.},
  \bibinfo{author}{Vercauteren, T.}, \bibinfo{author}{Zhang, S.},
  \bibinfo{year}{2020}.
\newblock \bibinfo{title}{{CA-Net}: Comprehensive attention convolutional
  neural networks for explainable medical image segmentation}.
\newblock \bibinfo{journal}{IEEE Transactions on Medical Imaging} .
\bibitem[{Hu et~al.(2016)Hu, Peng, Tai and Tang}]{hu2016network}
\bibinfo{author}{Hu, H.}, \bibinfo{author}{Peng, R.}, \bibinfo{author}{Tai,
  Y.W.}, \bibinfo{author}{Tang, C.K.}, \bibinfo{year}{2016}.
\newblock \bibinfo{title}{Network trimming: A data-driven neuron pruning
  approach towards efficient deep architectures}.
\newblock \bibinfo{journal}{arXiv preprint arXiv:1607.03250} .
\bibitem[{Hu et~al.(2018)Hu, Shen and Sun}]{hu2018squeeze}
\bibinfo{author}{Hu, J.}, \bibinfo{author}{Shen, L.}, \bibinfo{author}{Sun,
  G.}, \bibinfo{year}{2018}.
\newblock \bibinfo{title}{Squeeze-and-excitation networks}, in:
  \bibinfo{booktitle}{computer vision and pattern recognition}, pp.
  \bibinfo{pages}{7132--7141}.
\bibitem[{Ioffe and Szegedy(2015)}]{ioffe2015batch}
\bibinfo{author}{Ioffe, S.}, \bibinfo{author}{Szegedy, C.},
  \bibinfo{year}{2015}.
\newblock \bibinfo{title}{Batch normalization: Accelerating deep network
  training by reducing internal covariate shift}, in:
  \bibinfo{booktitle}{International Conference on Machine Learning}, pp.
  \bibinfo{pages}{448--456}.
\bibitem[{Isensee et~al.(2018)Isensee, Petersen, Klein, Zimmerer, Jaeger, Kohl,
  Wasserthal, Koehler, Norajitra, Wirkert et~al.}]{isensee2018nnu}
\bibinfo{author}{Isensee, F.}, \bibinfo{author}{Petersen, J.},
  \bibinfo{author}{Klein, A.}, \bibinfo{author}{Zimmerer, D.},
  \bibinfo{author}{Jaeger, P.F.}, \bibinfo{author}{Kohl, S.},
  \bibinfo{author}{Wasserthal, J.}, \bibinfo{author}{Koehler, G.},
  \bibinfo{author}{Norajitra, T.}, \bibinfo{author}{Wirkert, S.}, et~al.,
  \bibinfo{year}{2018}.
\newblock \bibinfo{title}{nnu-net: Self-adapting framework for u-net-based
  medical image segmentation}.
\newblock \bibinfo{journal}{arXiv preprint arXiv:1809.10486} .
\bibitem[{Lei et~al.(2020)Lei, Huang, Li, Li, Bian, Chou, Qin, Zhou, Gong and
  Cheng}]{lei2020self}
\bibinfo{author}{Lei, B.}, \bibinfo{author}{Huang, S.}, \bibinfo{author}{Li,
  H.}, \bibinfo{author}{Li, R.}, \bibinfo{author}{Bian, C.},
  \bibinfo{author}{Chou, Y.H.}, \bibinfo{author}{Qin, J.},
  \bibinfo{author}{Zhou, P.}, \bibinfo{author}{Gong, X.},
  \bibinfo{author}{Cheng, J.Z.}, \bibinfo{year}{2020}.
\newblock \bibinfo{title}{Self-co-attention neural network for anatomy
  segmentation in whole breast ultrasound}.
\newblock \bibinfo{journal}{Medical Image Analysis} , \bibinfo{pages}{101753}.
\bibitem[{Li et~al.(2020)Li, Yu, Chen, Fu, Xing and
  Heng}]{li2020transformation}
\bibinfo{author}{Li, X.}, \bibinfo{author}{Yu, L.}, \bibinfo{author}{Chen, H.},
  \bibinfo{author}{Fu, C.W.}, \bibinfo{author}{Xing, L.},
  \bibinfo{author}{Heng, P.A.}, \bibinfo{year}{2020}.
\newblock \bibinfo{title}{Transformation-consistent self-ensembling model for
  semisupervised medical image segmentation}.
\newblock \bibinfo{journal}{IEEE Transactions on Neural Networks and Learning
  Systems} \bibinfo{volume}{32}, \bibinfo{pages}{523--534}.
\bibitem[{Litjens et~al.(2012)Litjens, Debats, van~de Ven, Karssemeijer and
  Huisman}]{litjens2012pattern}
\bibinfo{author}{Litjens, G.}, \bibinfo{author}{Debats, O.},
  \bibinfo{author}{van~de Ven, W.}, \bibinfo{author}{Karssemeijer, N.},
  \bibinfo{author}{Huisman, H.}, \bibinfo{year}{2012}.
\newblock \bibinfo{title}{A pattern recognition approach to zonal segmentation
  of the prostate on {MRI}}, in: \bibinfo{booktitle}{International Conference
  on Medical Image Computing and Computer-Assisted Intervention},
  \bibinfo{organization}{Springer}. pp. \bibinfo{pages}{413--420}.
\bibitem[{Liu et~al.(2020)Liu, Kurgan, Wu and Wang}]{liu2020attention}
\bibinfo{author}{Liu, L.}, \bibinfo{author}{Kurgan, L.}, \bibinfo{author}{Wu,
  F.X.}, \bibinfo{author}{Wang, J.}, \bibinfo{year}{2020}.
\newblock \bibinfo{title}{Attention convolutional neural network for accurate
  segmentation and quantification of lesions in ischemic stroke disease}.
\newblock \bibinfo{journal}{Medical Image Analysis} \bibinfo{volume}{65},
  \bibinfo{pages}{101791}.
\bibitem[{Ma et~al.(2020)Ma, Wang, An, Ge, Yu, Chen, Zhu, Dong, He, He
  et~al.}]{ma2020towards}
\bibinfo{author}{Ma, J.}, \bibinfo{author}{Wang, Y.}, \bibinfo{author}{An, X.},
  \bibinfo{author}{Ge, C.}, \bibinfo{author}{Yu, Z.}, \bibinfo{author}{Chen,
  J.}, \bibinfo{author}{Zhu, Q.}, \bibinfo{author}{Dong, G.},
  \bibinfo{author}{He, J.}, \bibinfo{author}{He, Z.}, et~al.,
  \bibinfo{year}{2020}.
\newblock \bibinfo{title}{Towards efficient covid-19 ct annotation: A benchmark
  for lung and infection segmentation}.
\newblock \bibinfo{journal}{arXiv preprint arXiv:2004.12537} .
\bibitem[{Menze et~al.(2014)Menze, Jakab, Bauer, Kalpathy-Cramer, Farahani,
  Kirby, Burren, Porz, Slotboom, Wiest et~al.}]{menze2014multimodal}
\bibinfo{author}{Menze, B.H.}, \bibinfo{author}{Jakab, A.},
  \bibinfo{author}{Bauer, S.}, \bibinfo{author}{Kalpathy-Cramer, J.},
  \bibinfo{author}{Farahani, K.}, \bibinfo{author}{Kirby, J.},
  \bibinfo{author}{Burren, Y.}, \bibinfo{author}{Porz, N.},
  \bibinfo{author}{Slotboom, J.}, \bibinfo{author}{Wiest, R.}, et~al.,
  \bibinfo{year}{2014}.
\newblock \bibinfo{title}{The multimodal brain tumor image segmentation
  benchmark (brats)}.
\newblock \bibinfo{journal}{IEEE transactions on medical imaging}
  \bibinfo{volume}{34}, \bibinfo{pages}{1993--2024}.
\bibitem[{Milletari et~al.(2016)Milletari, Navab and Ahmadi}]{milletari2016v}
\bibinfo{author}{Milletari, F.}, \bibinfo{author}{Navab, N.},
  \bibinfo{author}{Ahmadi, S.A.}, \bibinfo{year}{2016}.
\newblock \bibinfo{title}{V-net: Fully convolutional neural networks for
  volumetric medical image segmentation}, in: \bibinfo{booktitle}{fourth
  international conference on 3D vision (3DV)}, pp. \bibinfo{pages}{565--571}.
\bibitem[{Molchanov et~al.(2016)Molchanov, Tyree, Karras, Aila and
  Kautz}]{molchanov2016pruning}
\bibinfo{author}{Molchanov, P.}, \bibinfo{author}{Tyree, S.},
  \bibinfo{author}{Karras, T.}, \bibinfo{author}{Aila, T.},
  \bibinfo{author}{Kautz, J.}, \bibinfo{year}{2016}.
\newblock \bibinfo{title}{Pruning convolutional neural networks for resource
  efficient inference}.
\newblock \bibinfo{journal}{arXiv preprint arXiv:1611.06440} .
\bibitem[{Mou et~al.(2020)Mou, Zhao, Fu, Liu, Cheng, Zheng, Su, Yang, Chen,
  Frangi et~al.}]{mou2020cs2}
\bibinfo{author}{Mou, L.}, \bibinfo{author}{Zhao, Y.}, \bibinfo{author}{Fu,
  H.}, \bibinfo{author}{Liu, Y.}, \bibinfo{author}{Cheng, J.},
  \bibinfo{author}{Zheng, Y.}, \bibinfo{author}{Su, P.}, \bibinfo{author}{Yang,
  J.}, \bibinfo{author}{Chen, L.}, \bibinfo{author}{Frangi, A.F.}, et~al.,
  \bibinfo{year}{2020}.
\newblock \bibinfo{title}{{CS2-Net}: Deep learning segmentation of curvilinear
  structures in medical imaging}.
\newblock \bibinfo{journal}{Medical Image Analysis} \bibinfo{volume}{67},
  \bibinfo{pages}{101874}.
\bibitem[{Nair et~al.(2020)Nair, Precup, Arnold and Arbel}]{nair2020exploring}
\bibinfo{author}{Nair, T.}, \bibinfo{author}{Precup, D.},
  \bibinfo{author}{Arnold, D.L.}, \bibinfo{author}{Arbel, T.},
  \bibinfo{year}{2020}.
\newblock \bibinfo{title}{Exploring uncertainty measures in deep networks for
  multiple sclerosis lesion detection and segmentation}.
\newblock \bibinfo{journal}{Medical image analysis} \bibinfo{volume}{59},
  \bibinfo{pages}{101557}.
\bibitem[{van Rijthoven et~al.(2020)van Rijthoven, Balkenhol, Silina, van~der
  Laak and Ciompi}]{van2020hooknet}
\bibinfo{author}{van Rijthoven, M.}, \bibinfo{author}{Balkenhol, M.},
  \bibinfo{author}{Silina, K.}, \bibinfo{author}{van~der Laak, J.},
  \bibinfo{author}{Ciompi, F.}, \bibinfo{year}{2020}.
\newblock \bibinfo{title}{Hooknet: multi-resolution convolutional neural
  networks for semantic segmentation in histopathology whole-slide images}.
\newblock \bibinfo{journal}{Medical Image Analysis} , \bibinfo{pages}{101890}.
\bibitem[{Ronneberger et~al.(2015)Ronneberger, Fischer and
  Brox}]{ronneberger2015u}
\bibinfo{author}{Ronneberger, O.}, \bibinfo{author}{Fischer, P.},
  \bibinfo{author}{Brox, T.}, \bibinfo{year}{2015}.
\newblock \bibinfo{title}{U-net: Convolutional networks for biomedical image
  segmentation}, in: \bibinfo{booktitle}{International Conference on Medical
  image computing and computer-assisted intervention}, pp.
  \bibinfo{pages}{234--241}.
\bibitem[{Schlemper et~al.(2019)Schlemper, Oktay, Schaap, Heinrich, Kainz,
  Glocker and Rueckert}]{schlemper2019attention}
\bibinfo{author}{Schlemper, J.}, \bibinfo{author}{Oktay, O.},
  \bibinfo{author}{Schaap, M.}, \bibinfo{author}{Heinrich, M.},
  \bibinfo{author}{Kainz, B.}, \bibinfo{author}{Glocker, B.},
  \bibinfo{author}{Rueckert, D.}, \bibinfo{year}{2019}.
\newblock \bibinfo{title}{Attention gated networks: Learning to leverage
  salient regions in medical images}.
\newblock \bibinfo{journal}{Medical image analysis} \bibinfo{volume}{53},
  \bibinfo{pages}{197--207}.
\bibitem[{Selvaraju et~al.(2017)Selvaraju, Cogswell, Das, Vedantam, Parikh and
  Batra}]{selvaraju2017grad}
\bibinfo{author}{Selvaraju, R.R.}, \bibinfo{author}{Cogswell, M.},
  \bibinfo{author}{Das, A.}, \bibinfo{author}{Vedantam, R.},
  \bibinfo{author}{Parikh, D.}, \bibinfo{author}{Batra, D.},
  \bibinfo{year}{2017}.
\newblock \bibinfo{title}{Grad-cam: Visual explanations from deep networks via
  gradient-based localization}, in: \bibinfo{booktitle}{international
  conference on computer vision}, pp. \bibinfo{pages}{618--626}.
\bibitem[{Simpson et~al.(2019)Simpson, Antonelli, Bakas, Bilello, Farahani,
  Van~Ginneken, Kopp-Schneider, Landman, Litjens, Menze
  et~al.}]{simpson2019large}
\bibinfo{author}{Simpson, A.L.}, \bibinfo{author}{Antonelli, M.},
  \bibinfo{author}{Bakas, S.}, \bibinfo{author}{Bilello, M.},
  \bibinfo{author}{Farahani, K.}, \bibinfo{author}{Van~Ginneken, B.},
  \bibinfo{author}{Kopp-Schneider, A.}, \bibinfo{author}{Landman, B.A.},
  \bibinfo{author}{Litjens, G.}, \bibinfo{author}{Menze, B.}, et~al.,
  \bibinfo{year}{2019}.
\newblock \bibinfo{title}{A large annotated medical image dataset for the
  development and evaluation of segmentation algorithms}.
\newblock \bibinfo{journal}{arXiv preprint arXiv:1902.09063} .
\bibitem[{Snell et~al.(2017)Snell, Swersky and Zemel}]{snell2017prototypical}
\bibinfo{author}{Snell, J.}, \bibinfo{author}{Swersky, K.},
  \bibinfo{author}{Zemel, R.}, \bibinfo{year}{2017}.
\newblock \bibinfo{title}{Prototypical networks for few-shot learning}, in:
  \bibinfo{booktitle}{Advances in neural information processing systems}, pp.
  \bibinfo{pages}{4077--4087}.
\bibitem[{Wang et~al.(2020)Wang, Cao, Ma, Zheng and Meng}]{wang2020pairwise}
\bibinfo{author}{Wang, R.}, \bibinfo{author}{Cao, S.}, \bibinfo{author}{Ma,
  K.}, \bibinfo{author}{Zheng, Y.}, \bibinfo{author}{Meng, D.},
  \bibinfo{year}{2020}.
\newblock \bibinfo{title}{Pairwise learning for medical image segmentation}.
\newblock \bibinfo{journal}{Medical Image Analysis} \bibinfo{volume}{67},
  \bibinfo{pages}{101876}.
\bibitem[{Zhou et~al.(2018)Zhou, Bau, Oliva and
  Torralba}]{zhou2018interpreting}
\bibinfo{author}{Zhou, B.}, \bibinfo{author}{Bau, D.}, \bibinfo{author}{Oliva,
  A.}, \bibinfo{author}{Torralba, A.}, \bibinfo{year}{2018}.
\newblock \bibinfo{title}{Interpreting deep visual representations via network
  dissection}.
\newblock \bibinfo{journal}{IEEE transactions on pattern analysis and machine
  intelligence} \bibinfo{volume}{41}, \bibinfo{pages}{2131--2145}.
\bibitem[{Zhou et~al.(2016)Zhou, Khosla, Lapedriza, Oliva and
  Torralba}]{zhou2016learning}
\bibinfo{author}{Zhou, B.}, \bibinfo{author}{Khosla, A.},
  \bibinfo{author}{Lapedriza, A.}, \bibinfo{author}{Oliva, A.},
  \bibinfo{author}{Torralba, A.}, \bibinfo{year}{2016}.
\newblock \bibinfo{title}{Learning deep features for discriminative
  localization}, in: \bibinfo{booktitle}{computer vision and pattern
  recognition}, pp. \bibinfo{pages}{2921--2929}.
\bibitem[{Zhou et~al.(2019)Zhou, Siddiquee, Tajbakhsh and
  Liang}]{zhou2019unetplusplus}
\bibinfo{author}{Zhou, Z.}, \bibinfo{author}{Siddiquee, M.M.R.},
  \bibinfo{author}{Tajbakhsh, N.}, \bibinfo{author}{Liang, J.},
  \bibinfo{year}{2019}.
\newblock \bibinfo{title}{Unet++: Redesigning skip connections to exploit
  multiscale features in image segmentation}.
\newblock \bibinfo{journal}{IEEE transactions on medical imaging}
  \bibinfo{volume}{39}, \bibinfo{pages}{1856--1867}.
\bibitem[{Zhu and Gupta(2017)}]{zhu2017prune}
\bibinfo{author}{Zhu, M.}, \bibinfo{author}{Gupta, S.}, \bibinfo{year}{2017}.
\newblock \bibinfo{title}{To prune, or not to prune: exploring the efficacy of
  pruning for model compression}.
\newblock \bibinfo{journal}{arXiv preprint arXiv:1710.01878} .

\end{thebibliography}

\end{document}